\documentclass[aps,pre,reprint,longbibliography]{revtex4-1}
\usepackage{graphicx}
\usepackage{amsmath}
\usepackage{amssymb}
\usepackage{natbib}
\usepackage[usenames]{color}

\newcommand{\change}{\textcolor{black}}

\begin{document}

\title{ 
Thermodynamic Stability of Driven Open Systems and \\
Control of Phase Separation by Electro-autocatalysis
}

\author{Martin Z. Bazant 
}

\affiliation{Departments of Chemical Engineering and Mathematics,
  Massachusetts Institute of Technology, Cambridge, MA 02139 USA}
\affiliation{ \protect{\em Present address:} Department of Materials Science and Engineering and SUNCAT Interfacial Science and Catalysis, Stanford University, Stanford, CA 94305}
\setcounter{page}{1} \date{\today}
\label{firstpage}

\begin{abstract}
Motivated by the possibility of electrochemical control of phase separation, a variational theory
of thermodynamic stability is developed for driven reactive mixtures, based on a nonlinear generalization
of \change{the} Cahn-Hilliard and Allen-Cahn equations. The Glansdorff-Prigogine stability criterion
is extended for driving chemical work, based on variations of nonequilibrium Gibbs free energy.
Linear stability is generally determined by the competition of chemical diffusion and driven autocatalysis.
Novel features arise for electrochemical systems, related to controlled total current (galvanostatic
operation), concentration-dependent exchange current (Butler-Volmer kinetics), and
negative differential reaction resistance (Marcus kinetics). The theory shows how spinodal decomposition
can be controlled by solo-autocatalytic charge transfer, with only a single Faradaic
reaction. Experimental evidence is presented for intercalation and electrodeposition in rechargeable
batteries, and further applications are discussed in solid state ionics, electrovariable optics,
electrochemical precipitation, and biological pattern formation.
\end{abstract}

\maketitle

\section{ Introduction }

This Faraday Discussion\footnote{ This invited paper will be published in a special issue of {\it Faraday Discussions} for Chemical Physics of Electroactive Materials, April 10-12, 2017, Cambridge, UK. } focuses on the use of electric fields to control the dynamical response of materials, such as electroactuation of polymer gels and electrovariable optics with plasmonic nanoparticles. Although it has not been widely recognized, these phenomena could be strongly affected by phase separation of the constituents into domains of different density \change{or} chemical identity. 
Here we consider the possibility of controlling such phase separation by electrochemical reactions. This raises 
 fundamental questions about thermodynamic stability, which we motivate by first summarizing the physical picture behind our results.

\section{ Physical Picture }
\label{sec:phys}

\subsection{ Thermodynamic Stability Near Equilibrium }

Consider a system containing a chemical species A at uniform concentration $c$, which is thermodynamically unstable to concentration fluctuations.  In particular, attractive inter-particle forces favor phase separation into stables phases of higher and lower concentration, which correspond to local minima of the homogeneous Gibbs  free energy $g_h(c)$.  As discussed below, Gibbs himself developed the original stability criterion for chemical mixture near equilibrium:
\begin{equation}
\mbox{ Stable:} \ \ \frac{d^2 g_h}{dc^2} = \frac{d\mu_h}{dc} > 0   \label{eq:gibbs}
\end{equation}
where $\mu_h(c)$ is the diffusional chemical potential of the homogeneous mixture, defined as the change in free energy upon adding a particle of species A  at constant temperature and pressure.

\begin{figure}
\begin{center}
\includegraphics[width=2.3in]{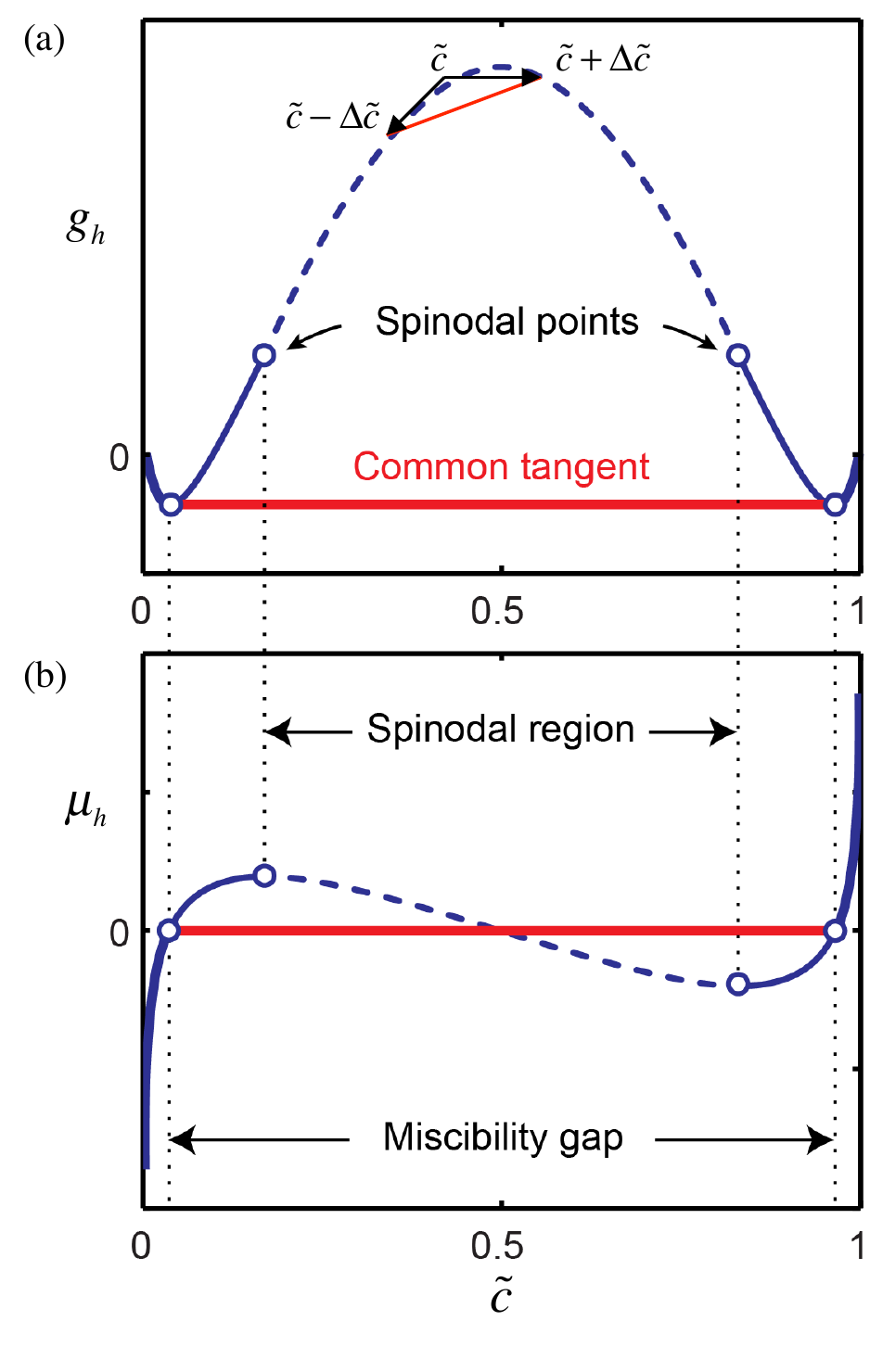}
\caption{ Thermodynamic stability an inert, homogeneous binary mixture (described by the regular solution model~\cite{cahn1958,kom,bazant2013}).  (a) Homogeneous free energy and (b) diffusional chemical potential versus dimensionless concentration, showing the common tangent construction for phase separation in the miscibility gap (red). The secant construction for linear instability in the chemical spinodal region (dashed blue) is shown in (a).   \label{fig:spin}  }
\end{center}
\end{figure}

The Gibbs criterion (\ref{eq:gibbs}) has a simple graphical interpretation, shown in Fig. ~\ref{fig:spin} for a binary mixture with two stable equilibrium states, corresponding to two local minima of $g_h(c)$ or zeros of $\mu_h(c)=g_h^\prime(c)$. In the ``miscibility gap" between the minima, it is favorable to phase separate into a linear combination of the two stable states having the same average concentration, whose free energy lies on a common tangent construction.  The same principle can be applied to small concentration fluctuations using a local secant construction, which shows that stable concentrations correspond to a locally convex free energy, $g_h^{\prime\prime}(c)>0$, or increasing chemical potential, $\mu_h^\prime(c)>0$. Within the ``chemical spinodal" where convexity is lost, $g_h^{\prime\prime}(c)=\mu_h^\prime(c)<0$, the system is unstable to spontaneous phase separation (``spinodal decomposition")~\cite{kom}.

\subsection{ Stability of Mixtures with Driven Chemical Reactions }

The theory of thermodynamic stability has been extended to include chemical reaction networks in closed bulk systems with porous boundaries~\cite{kondepudi_book}, such as biological cells, but here we focus instead on driven chemical reactions in open bulk systems.  The basic principles are illustrated by driven adsorption,
\begin{equation}
\mbox{M}_{res} \longrightarrow \mbox{M}  \label{eq:A}
\end{equation}
where a single species M evolves with local chemical potential $\mu(x,t)$ and undergoes homogeneous reactions with a reservoir at constant chemical potential, $\mu_{res}$, where it takes the form of (possibly different) species $M_{res}$.   For bulk mixtures, this model could describe a reactive species $M$ at low concentration in a  sea of  equilibrated molecules, which includes the reaction product $M_{res}$, as in open-system models of self-organization in biological cells~\cite{popovic2014entropy}.  

\begin{figure}[tb]
\begin{center}
\includegraphics[width=3in]{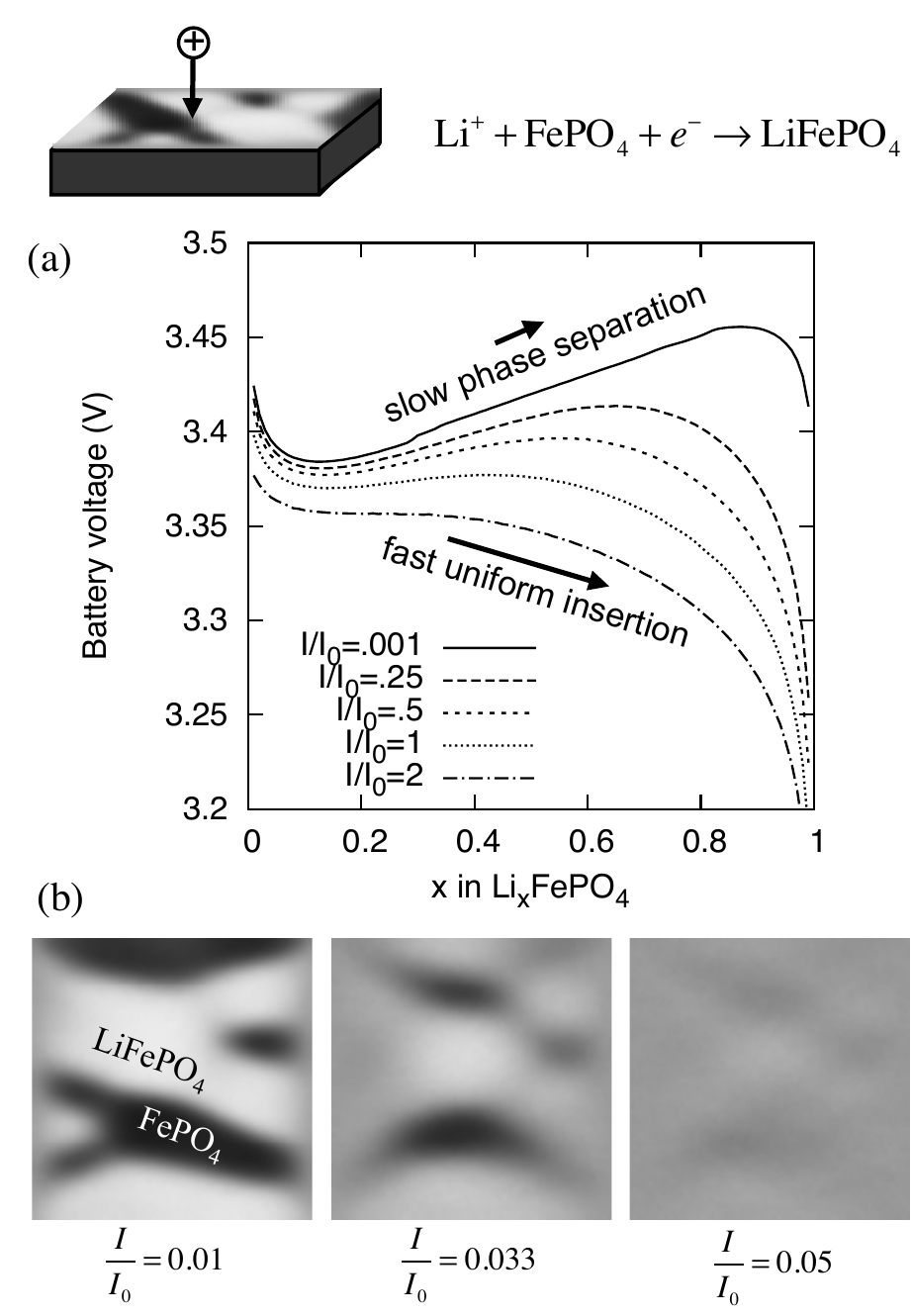}
\caption{ Control of coherent phase separation in a binary solid Li-ion battery cathode (Li$_X$FePO$_4$) by Faradaic insertion reactions.  (a) Predicted battery voltage versus lithium metal ($V = V^\Theta - \mu/e$) and (b) surface lithium concentration profiles at $X=0.6$ for different applied currents, scaled to a reference exchange current, $I_0$. [Adapted from  \citet{cogswell2012}] \label{fig:cogswell}  }
\end{center}
\end{figure}

The same model also describes a wide variety of adsorption phenomena at solid or liquid interfaces, such as monolayer adsorption, where attractive lateral forces can drive pattern formation~\cite{bazant2013}. This tendency for clustering modifies the classical theory of surface adsorption~\cite{seri1993brunauer,nikitas1996simple} and sorption hysteresis in porous media~\cite{bazant2012theory}.  Similar phenomena can occur for the solid-state insertion of bulk neutral species, such as hydrogen into palladium hydride~\cite{baldi2014situ,tang2011observations,ulvestad2015avalanching}, or charged species at electrodes, such as lithium ions into iron phosphate ~\cite{bazant2013,bai2011,cogswell2012,lim2016origin}, shown in Fig. ~\ref{fig:cogswell}.

A key result of our general stability analysis below is that a fast driven reaction can suppress phase separation at constant potential $\mu_{res}$ if the reaction rate $R$ decreases with reaction extent,
\begin{equation}
\mbox{Stable:} \ \ \ \left(\frac{dR}{dc}\right)_{\mu_{res}} < 0 \ \ \ \mbox{(constant potential)}    \label{eq:Rstab}
\end{equation}
Electrochemical systems offer the unique capability of controlling the rate of Faradaic relations, and this leads to a new phenomena of phase separation at constant current. 
In the usual case of positive reaction resistance (defined below), phase separation is suppressed if the reservoir potential increases with reaction extent:
\begin{equation}
\mbox{Stable:} \ \ \ \left(\frac{d\mu_{res}}{dc}\right)_{R} > 0 \ \ \ \mbox{(constant current)}    \label{eq:murstab}
\end{equation}
which is a generalization of the Gibbs criterion (\ref{eq:gibbs}) for a chemically driven, open system.   This effect is clearly seen in the lithium insertion simulations of Fig. \ref{fig:cogswell}, where the battery voltage becomes monotonically decreasing with concentration ($\frac{d\mu}{dc} = - e\frac{dV}{dc} > 0$), as concentration fluctuations disappear above a critical current.  

In summary, {\it phase separation is reduced if the reaction is auto-inhibitory} (either slows down or becomes harder to drive), {\it or enhanced if it is auto-catalytic} (either accelerates or becomes easier to drive). 

\subsection{ Solo-autocatalysis }

We refer to this nonlinearity for a single reaction in a concentrated mixture as ``solo-autocatalysis" to distinguish it from the traditional concept of ``collective autocatalysis" for chemical reaction networks in dilute mixtures, governed by mass action kinetics.  Solo-autocatalysis is an inescapable feature of adsorption, intercalation and deposition reactions.
Whenever the product (or reactant) occupies a finite set of sites, it necessarily affects the subsequent reaction rate.  Adsorption reactions are typically  solo-autoinhibitory (rate suppressing) at high concentration, as product covers the active sites.  Since the reaction creates a particle M while destroying a vacancy V, the vacancy can be viewed as an adsorption catalyst, $\mbox{M}_{res} + \mbox{V} \longrightarrow \mbox{M}$, which slowly disappears as the reaction progresses.
Vacancies can also be viewed as a distinct chemical species in a reactive binary mixture with the adsorbed particles.  The total volume constraint yields a single concentration variable, $c_M$, $c_V$, or dimensionless coverage, $\tilde{c} = c_M/c_s$ (where $c_s=$ site concentration),  which evolves in response to differences in ``diffusional chemical potential", $\mu = \mu_M - \mu_V$, either by diffusion or reactions~\cite{kom,bazant2013,nauman2001,carati1997chemical}.  The same applies to the isomerization reaction, $\mbox{M}\longrightarrow\mbox{V}$, in a closed system~\cite{lamorgese2016spinodal}, which corresponds to $\mu_{res}=0$.

In general, it may not be possible to identify vacancies or other catalytic species, and yet the reaction rate still depends on concentration.   In particular, electrochemical reactions tend to be solo-autocatalytic (rate enhancing) at low concentration, as redox active molecules increase the exchange rate for electron transfer~\cite{bard_book,bazant2013}, while remaining auto-inhibitory at high concentration.  The result is a ``volcano" shaped exchange current versus concentration, which is usually assumed to be symmetric,  $I_0 \sim \sqrt{\tilde{c}(1-\tilde{c})}$,
in models of Li-ion batteries~\cite{newman_book,doyle1993} and fuel cells~\cite{kulikovsky_book,eikerling1998,fu2015heterogeneous}.  
In contrast, the theory of charge transfer based on non-equilibrium thermodynamics predicts an asymmetric exchange-current volcano favoring higher rates at low concentrations, considering only site exclusion in the transition state~\cite{bazant2013}.  As we shall see, this  turns out to be the key property that enables the control of phase separation~\cite{bai2011,cogswell2012,lim2016origin}.

\subsection{ Control of Phase Separation by Electro-autocatalysis }

\begin{figure*}
\begin{center}
\includegraphics[width=5in]{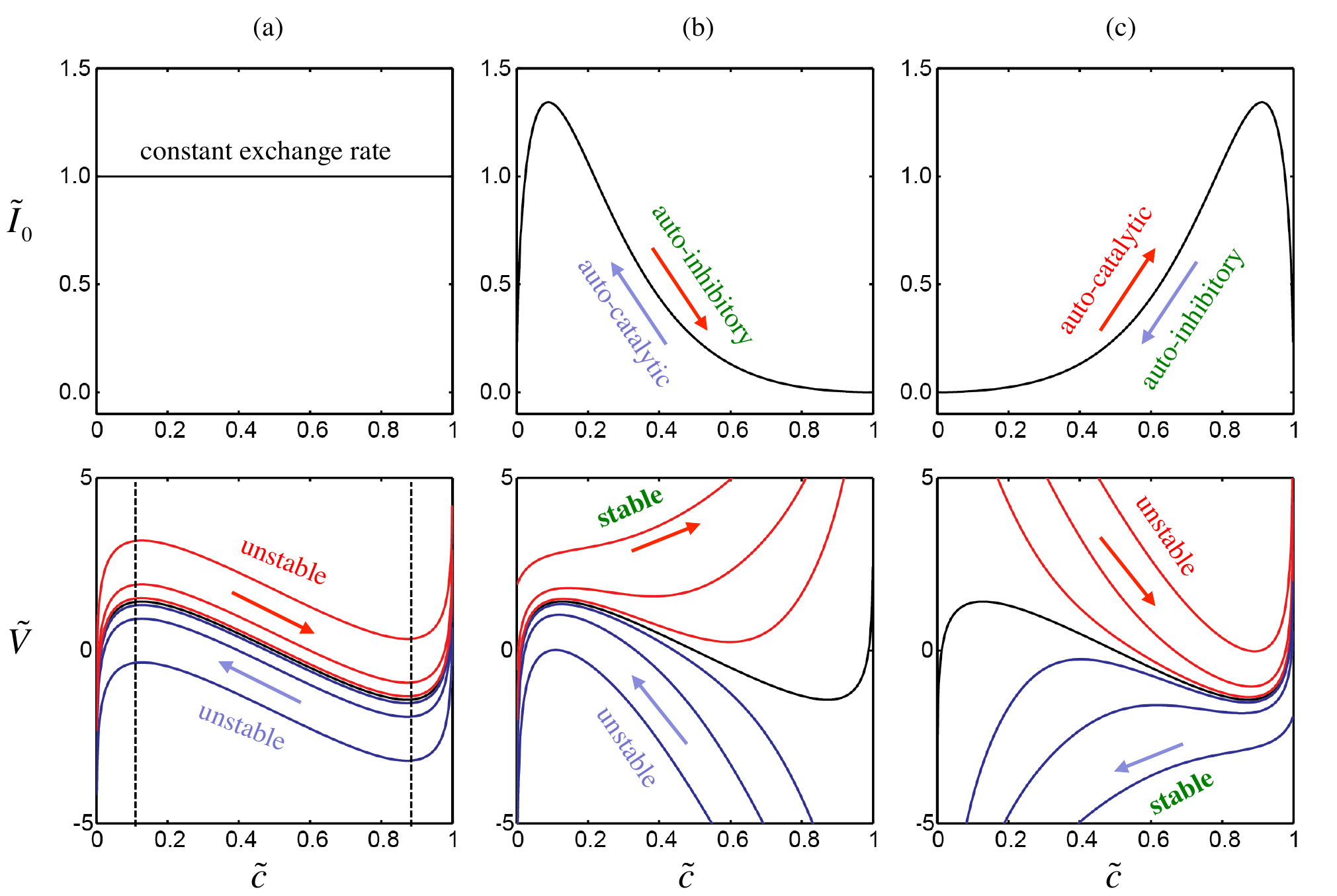}
\end{center}
\caption{  Principles of thermodynamic stability controlled by electro-autocatalysis. Top row: Dimensionless exchange current vs. product concentration $\tilde{I}_0(\tilde{c})$.  Bottom row: Dimensionless electrode voltage versus concentration at different applied currents for  insertion (red) and extraction (blue), where signs are chosen for anodic cation insertion to resemble neutral-species adsorption ($\tilde{V} = \tilde{\mu}_{res}$).  (a)  A non-autocatalytic reaction ($\tilde{I}_0^\prime=0$) simply shifts the potential curves up and down by constant activation potential, and thus cannot alter the spinodal region of instability (negative slope, between dashed lines).  (b) An auto-inhibitory reaction ($\tilde{I}_0^\prime<0$) in the spinodal reaction can suppress the instability (positive slope) leading to ``electrochemical freezing" above a critical insertion current, while further destabilizing the system during extraction.  Outside the spinodal, the reaction creates instability and leads to ``electrochemical melting" above a critical current, while further stabilizing the mixture during extraction.  (c) An auto-catalytic reaction in the spinodal region ($\tilde{I}_0^\prime>0$) has the opposite effect of destabilization during insertion and stabilization during extraction.
   \label{fig:I0} }
\end{figure*}
    
The fundamental mechanism for control of phase separation by driven autocatalysis is illustrated in Fig. ~\ref{fig:I0}, in the case of anodic ion insertion, or adsorption of a neutral species, at constant current.    The externally controlled potential $\mu_{res}$ is equal to the internal potential $\mu_h(c)$ (for a homogeneous base state) plus the affinity, $A=\mu_{res}-\mu_h(c)$, which controls the reaction rate.  In the case of Faradaic reactions transferring $n$ electrons, it is the (anodic) {\it activation overpotential}, $\eta = A/ne$, that controls the Faradaic (oxidation) current,  $I = nev$.  The simplest autocatalytic model has a separable form, $I = I_0(c) f(ne\eta/k_BT)$, with a concentration-dependent exchange current, $I_0(c)$ and monotonic overpotential dependence ($f^\prime>0$, $f^\prime(0)=1$, $f(0)=0$), as in the Bulter-Volmer equation~\cite{bard_book} and various generalizations for concentrated solutions~\cite{bazant2013}, considered below. 

The reservoir chemical potential, or cell voltage $V$, thus depends on concentration and the applied current, 
\begin{equation}
\tilde{\mu}_{res} = \tilde{V}-\tilde{V}_0 = \tilde{\mu}_h(c) +  f^{-1}\left(\frac{I}{I_0(c)}\right) \sim \tilde{\mu}_h(c) + \frac{I}{I_0(c)}
\end{equation}
where potential is scaled to $k_BT$ and voltage to $k_BT/ne$, $V_0$ is the open circuit voltage at $\mu=0$, and, for clarity, we linearize the overpotential dependence -- but not the autocatalytic concentration dependence.
As shown in Fig. \ref{fig:I0}(a)), for a non-autocatalytic reaction ($I_0^\prime=0$), the activation overpotential is constant, so the shape of the voltage profile and stability of the system cannot be altered by the reaction. 

Autocatalysis is required to alter thermodynamic stability.
As shown in Fig. \ref{fig:I0}(b)), for concentrations where the insertion reaction is auto-inhibitory ($I_0^\prime(c)<0$), the homogeneous state becomes stable ($\mu_{res}^\prime >0$) above a critical current, even within the spinodal region, which amounts to {\it electrochemical freezing} of a thermodynamically unstable mixture in a disordered state. The system's entropy is increased above its equilibrium value by applying external work to drive the reaction. This phenomenon is different from rapid quenching of a liquid to a metastable glass or amorphous solid, because the electrochemically frozen mixture is stable under the applied current. As soon as the current drops below the critical value, however, spontaneous phase separation occurs.  

Interestingly, when the current is reversed, the opposite phenomena occur. Phase separation is enhanced in the spinodal region, and the homogeneous mixture outside the spinodal  can be destabilized. The latter corresponds to {\it electrochemical melting} of a thermodynamically stable disordered state to form two ordered phases. Again this is not a transient phenomenon, but  a change of thermodynamic stability in which the external work driving the reaction makes it favorable to lower of the system's entropy.

As shown in Fig. \ref{fig:I0}(c)), for concentrations where the reaction is autocatalytic ($I_0^\prime(c)>0$), the system becomes more unstable with increasing insertion current. Above a critical insertion current, phase separation can occur outside the spinodal region, which corresponds to {\it electrochemical melting} of a thermodynamically stable mixture. Conversely, extraction currents now stabilize the system and can lead to electrochemical freezing of the spindoal region below a threshold negative current. 

In summary, the theory predicts the following effects of electro-autocatalysis on phase separation at constant current:
\begin{itemize}
\item  During periods of auto-inhibition ($I_0^\prime(c)<0$), the forward reaction ($I>0$) suppresses phase separation (completely for $I>I_c$), while the backward reaction ($I<0$) enhances it.
\item During periods of autocatalysis ($I_0^\prime>0$), the forward reaction ($I>0$) enhances phase separation, while the backward reaction ($I<0$) suppresses it.
\end{itemize}
These predictions have recently been verified in experiments on Li-ion battery materials, as discussed below in Section ~\ref{sec:expt}.

\subsection{ Nonequilibrium Gibbs Free Energy }

In the examples above,  the applied current appears to act as an independent state variable, analogous to temperature, pressure and concentration.  In hindsight, the reason is that constant current contributes a well-defined state-dependent excess energy (the activation overpotential) to the total {\it non-equilibrium} Gibbs free energy, $\mathcal{G}(c,I)$ of the driven open system.  Comparing Eqs. (\ref{eq:gibbs}) and (\ref{eq:gstab}), such a state function could be defined as 
\begin{equation}
\Delta\mathcal{G}(c,I) = \int_{c_0}^c \mu_{res}(c,I)\, dc =   \Delta\mathcal{G}_{eq}(c) + \Delta_i W_d(c,I)  \label{eq:dg1}
\end{equation}
where we define the reversible change in equilibrium free energy, associated with the transformation at zero current,
\begin{equation}
\Delta \mathcal{G}_{eq}(c)=\Delta\mathcal{G}(c,0) = g_h(c)-g_h(c_0)
\end{equation}
and the irreversible driving work done on the system at finite current,
\begin{equation} 
\Delta_i W_d(c,I) = \int_{c_0}^c A(c,I)\, dc = \int_{t_0}^t I^2\mathcal{R}_F(c,I)dt   \label{eq:wd1}
\end{equation}
For Faradaic reactions, the latter is equal to the time-integral of the electrical power, $P_e = I^2 \mathcal{R}_F$, where $\mathcal{R}_F = -\eta/I>0$ is the Faradaic resistance.  This simple example will help us generalize the theory of thermodynamic stability for driven open systems.


\subsection{ Driven Autocatalysis versus Chemical Diffusion }

The preceding simple analysis considers driven autocatalytic reactions which are fast compared with diffusion (large Damk\"ohler number, $\mbox{Da}>1$, defined below).  In the  opposite limit of negligible reactions, Cahn pioneered the theory of diffusion-driven spinodal decomposition~\cite{cahn1961,cahn1962sd,cahn1965phase}. The instability is controlled by the chemical diffusivity~\cite{kom},
\begin{equation}
\bar{\mathcal{D}} = \frac{ D\, c }{k_B T} \frac{d\mu_h}{dc}
\end{equation}
which enters Fick's law (flux $=-\bar{\mathcal{D}}\nabla c$) for a concentrated solution, where $D>0$ is the tracer diffusivity in the dilute limit~\cite{bazant2013}.  Outside the spinodal,  ``forward diffusion" ($\bar{\mathcal{D}}>0$) leads to familiar smooth concentration profiles, but   inside the spinodal, the system is destabilized by ``backward diffusion" ($\bar{\mathcal{D}}<0$) leading to phase separation.

Here, we show that thermodynamic stability of reactive mixtures is determined by the competition of autocatalysis and chemical diffusion. In driven open systems, such as electrochemical interfaces, this competition can be controlled by applied potentials and currents.  The theory predicts that stable equilibrium mixtures can be driven to form desired patterns by  electrochemical melting, while unstable mixtures can be driven to remain homogeneous by electrochemical freezing.  
These surprising phenomena not only have applications to electroactuation, but they also raise profound questions about nonequilibrium thermodynamics, as we now explain.

\section{ Background }
\label{sec:back}

In order to analyze the stability of driven open systems, we must first extend nonequilibrium chemical thermodynamics~\cite{kondepudi_book,degroot_book} for inhomogeneous systems, as described by phase-field models~\cite{kom,chen2002phase}, using the calculus of variations~\cite{gelfand_book}.
 
\subsection{  Gibbs' Stability Theory for Inert Mixtures }

Gibbs pioneered the theory of thermodynamic stability~\cite{gibbs_works}, based on the notion that entropy is maximized in equilibrium~\cite{kondepudi_book}. As such, any perturbation of a stable equilibrium must lower its entropy (or increase its free energy) according to
$
\Delta S = S - S_{eq} = \delta S + \frac{1}{2}\delta^2S+\ldots  
$, where the first and second variations of the entropy functional with respect to spatial perturbations in concentration, temperature, etc. must satisfy
\begin{equation}
\mbox{Stable equilibrium:} \ \ \delta S = 0 \mbox{ and } \delta^2 S < 0.   \label{eq:genstab}
\end{equation}
For fluctuations in temperature or volume, the Gibbs stability criterion implies positive heat capacity, $C_v>0$, and isothermal compressibility, $\kappa_T>0$.   

For concentration fluctuations $\{ \delta c_i \}$ at constant internal energy and volume, stable equilibrium requires~\cite{kondepudi_book}
\begin{equation}
\delta^2 S = - \int_V \sum_{i,j} \delta c_i\left(\frac{\delta}{\delta c_j} \frac{\mu_i}{T}\right)  \delta c_j dV < 0  \label{eq:mustab}
\end{equation}
where we define the (diffusional) chemical potential
\begin{equation}
\mu_i = \frac{\delta G}{\delta c_i}  \label{eq:dG1} 
\end{equation} 
as the first variational derivative of the Gibbs free energy with respect to the concentration of species $i$. This is the continuum analog of the familiar definition from statistical mechanics, $\mu_i = \left(\frac{\Delta G}{\Delta N_i}\right)_{T,P}$,  as the change in free energy from adding a particle of species $i$, where a ``particle" corresponds to a Dirac delta function added to the concentration profile at a given position~\cite{bazant2013}.
 
With this generalization, Gibbs'  maximum entropy condition, Eq. (\ref{eq:mustab}), implies that the  Hessian tensor of second variational derivatives, $G^{\prime\prime}$, must be positive definite in equilibrium, 
\begin{equation}
\frac{\delta\mu_i}{\delta c_j} = \frac{\delta^2G}{\delta c_i \delta c_j} = G^{\prime\prime}_{ij} > 0   \label{eq:dG2}
\end{equation}
(We write $T_{ij}>0$ if $\sum_{ij}\int_V  \delta u_i T_{ij}  \delta u_j\, dV > 0$ for all $\delta u_i$, $\delta u_j$.)
In the limit of long-wavelength fluctuations in a uniform system, this asserts that the homogeneous free energy density, $g_h(\{ c_i \})$, has a positive definite Hessian matrix of second partial concentration derivatives,
\begin{equation}
\bar{G}^{\prime\prime}_{ij} = \frac{\partial^2 g_h}{\partial c_i\, \partial c_j} > 0 \label{eq:gstab}
\end{equation}
In order words, in stable equilibrium, the free energy must be locally convex with respect to concentration, as shown in Fig.~\ref{fig:spin}.  The variational formula, Eq. (\ref{eq:dG2}), extends this concept to nonuniform systems and arises naturally in our nonequilibrium stability analysis below.

\subsection{ Thermodynamics of Inhomogeneous Systems }

In contrast to classical thermodynamic models~\cite{kondepudi_book}, we allow the Gibbs free energy functional, $G[\{c_i\}]$, to have explicit dependence on concentration gradients, which could arise from interfacial tension, elastic coherency strain, electrostatic energy, or other non-idealities of inhomogeneous systems.  In Eqs. (\ref{eq:dG1}) and (\ref{eq:dG2}), we introduce notation for the first, second, and higher variational derivatives,  
\begin{eqnarray}
\Delta G &=& \delta G + \frac{1}{2}\delta^2 G + \ldots  \\
&=& \int_V \sum_i \delta c_i  \left( \frac{\delta G}{\delta c_i} + \sum_j \delta c_j  \left( \frac{1}{2} \frac{\delta^2 G}{\delta c_i\delta c_j} + \ldots \right)\right) dV \nonumber   \label{eq:dGdef}
\end{eqnarray}
defined by the expansion of the free energy change in response to one, two or more simultaneous bulk concentration fluctuations (which vanish on the boundary), respectively. 

In order to describe the dynamics of phase separation, it is necessary to model interfacial tension between phases without artificially introducing sharp phase boundaries.  In 1893, Van der Waals first proposed adding a quadratic gradient penalty to the homogeneous free energy~\cite{van1893verhandel,widom1999we},
 \begin{equation}
G[c] = \int_V \left(\mu^\Theta c + g_h(c) +  \frac{K}{2}|\nabla c|^2 \right) dV  \label{eq:GCH1}
\end{equation}
where we include a reference chemical potential~\cite{bazant2013}, $\mu^\Theta$. The gradient penalty term, $\frac{K}{2}|\nabla c|^2 = \frac{\kappa}{2}|\nabla\tilde{c}|^2$,  is often written in terms of filling fraction, $\tilde{c}=c/c_s$, over sites of density $c_s$, where $\kappa=K c_s^2$, and can be adjusted to fit the tension and thickness of phase boundaries.  This visionary idea was somehow forgotten for over half a century, until its rediscovery in physics by Landau and Ginzburg~\cite{landau1950theory} (to describe magnetic flux in type II superconductors) and in materials science by Cahn and Hilliard~\cite{cahn1958} (to describe phase separation in solid binary alloys).  

Led by Cahn~\cite{cahn1959-1,cahn1959-2,cahn1961,cahn1962sd,cahn1962cohnuc,allen1972ground,cahn1977critical}, this approach paved the way for modern phase-field models~\cite{chen2002phase,kom}, which approximate phase boundaries as localized, but continuous, ``diffuse interfaces". Taking a functional derivative of Eq. (\ref{eq:GCH}), the diffusional chemical potential (per site) $\mu$ and its homogeneous limit $\mu_h$  are given by
\begin{equation}
\mu = \mu_h -  K \nabla^2 c \ \mbox{ and } \  
\mu_h = \mu^\Theta + \frac{dg_h}{dc}.   \label{eq:mu1}
\end{equation}
Equilibrium concentration profiles satisfy the Beltrami equation, $\mu =\frac{\delta G}{\delta c}=$ constant.  Solutions in the miscibility gap describe uniform stable domains separated by diffuse phase boundaries, whose width, $\lambda = \sqrt{\kappa/c_s\Omega}$, and interfacial tension, $\gamma = \sqrt{\kappa c_s \Omega}$, are related to the gradient penalty $\kappa$ and a characteristic  energy barrier between the stable concentrations, $\Omega$, e.g. the regular solution parameter for pairwise interatomic forces~\cite{cahn1958,kom}.

For multicomponent, anisotropic, inhomogeneous systems, the Cahn-Hilliard free energy, chemical potentials, and Hessian tensor are given by
\begin{eqnarray}
G &=& \int_V \left( \sum_i \mu^\Theta_i c_i  + g_h(\{c_i\}) + \frac{1}{2}\sum_{ij}\nabla c_i \cdot K_{ij} \nabla c_j \right) dV  \label{eq:GCH} \\
\mu_i &=& \mu^{\Theta}_i + \frac{\partial g_h}{\partial c_i} - \sum_j \nabla\cdot K_{ij} \nabla c_j \\
\frac{\delta\mu_i}{\delta c_j} &=& \frac{\partial^2 g_h}{\partial c_i\partial c_j} 
+ \frac{\nabla\delta c_i}{\delta c_i} \cdot K_{ij} \frac{\nabla\delta c_j}{\delta c_j}   \label{eq:d2Gij} 
\end{eqnarray}
where the Hessian depends on gradients of the fluctuations, according to Eq. (\ref{eq:dGdef}). 

\subsection{ Linear Irreversible Thermodynamics of Diffusion }

Gradients in chemical potential provide thermodynamic forces that drive diffusional fluxes, respectively, 
\begin{equation}
F_i = - \nabla \frac{\mu_i}{T} \ \mbox{ and } \ J_i = \sum_j L_{ij} F_j   \label{eq:LITflux}
\end{equation}
where we make the ubiquitous approximation of Linear Irreversible Thermodynamics (LIT)~\cite{kondepudi_book}, which is valid close to local equilibrium. The linear  response matrix must be symmetric, $L_{ij}=L_{ji}$ (Onsager relation), and positive definite, in order to ensure a positive entropy production rate by diffusion,
\begin{equation}
\frac{d_i S}{dt} = \int_V \left(\sum_i F_i J_i\right) dV = \int_V \left(\sum_{ij} F_i L_{ij} F_j\right) dV > 0
\end{equation}
Mass conservation with LIT fluxes yields the (multi-component) Cahn-Hilliard equation,
\begin{equation}
\frac{\partial c_i}{\partial t} =  \nabla\cdot \sum_j L_{ij} \nabla \frac{\delta G}{\delta c_j}  \label{eq:CHeq}
\end{equation}
which is the standard model for phase separation by diffusion in a closed system~\cite{chen2002phase,kom}, including linear instability and spinodal decomposition~\cite{cahn1961,cahn1962sd,cahn1962cohnuc}.
The Onsager coefficients are related to the mobility tensor (drift velocity per force) via $L_{ij} = M_{ij} c_j$.
For a single diffusing species, the tracer diffusivity satisfies the Einstein relation, $D(c) = M(c) k_BT$, and takes the form $D = D_0(1-\tilde{c})$ or $L\sim \tilde{c}(1-\tilde{c})$ in a binary mixture~\cite{nauman2001}, to reflect the crowding of sites~\cite{bazant2013}. 

The phase-field LIT formalism can be extended to electrochemical systems~\cite{bazant2013}, which have long-range Coulomb forces in addition to the short-range forces that determine $g_h$. The electrochemical potential is defined by adding the electrostatic energy $q_i \phi$ to $\mu$, and the associated Nernst-Planck LIT flux (ionic current) includes contributions from diffusion and electromigration. The mobility matrix $L_{ij}$ is usually assumed to be diagonal, but this neglects strongly coupled fluxes at high concentrations, where strong Coulomb correlations may yield negative off-diagonal coefficients~\cite{gavish2016theory}. The electrostatic potential of mean force, $\phi$, is determined either by electroneutrality or Poisson's equation. 

\subsection{ Prigogine's Stability Theory for Reactive Mixtures }

Let us now consider the effect of chemical reactions, $\mbox{M}_{r,m} = \sum_i s_{r,m,i} \mbox{M}_{r,m,i} \longleftrightarrow \sum_j s_{p,m,j} \mbox{M}_{p,m,j} = \mbox{M}_{p,m}$, where $M_{r,m}$ and $M_{p,m}$ are the reactant and product complexes of the $m$th reaction with total chemical potentials,  $\mu_{r,m}= \sum_i s_{r,m,i} \mu_{r,m,i} \ \mbox{ and } \ \mu_{p,m}= \sum_j s_{p,m,j} \mu_{p,m,j},$
and stoichiometric coefficients, $\{s_{r,m,i}\}$ and $\{s_{p,m,j}\}$, respectively.  For electrochemical reactions, the chemical species $\{M_i\}$ include both ions and electrons.  The thermodynamic driving force for a reaction is the change in Gibbs free energy~\cite{prigogine_book,kondepudi_book}, 
$\Delta_r G_m = \mu_{p,m} - \mu_{r,m}$, which is equal to the difference in diffusional chemical potentials~\cite{bazant2013}.
For a Faradaic reduction reaction transferring $n$ electrons, the activation overpotential,  $\eta_m = \Delta_r G_m/ne$, 
is the free energy of the net reduction reaction per charge~\cite{bazant2013}.

De Donder pioneered non-equilibrium chemical thermodynamics and related the free energy of reaction to the chemical affinity~\cite{prigogine1980nonequilibrium,de1936affinite}, 
\begin{equation}
A_m = - \left( \frac{\partial G}{\partial \xi_m} \right)_{T,P} =  -\Delta_r G_m 
\end{equation} 
where $G$ is the total  Gibbs free energy, including reactants and products, and $\xi_m$ is the extent of the reaction.  
He also argued that the free energy of reaction contributes to Clausius' ``uncompensated heat", $dQ'$ (or irreversible entropy production, $d_iS$, in modern terminology~\cite{prigogine1980nonequilibrium,prigogine1978coupling}) and introduced the equivalent definition,
\begin{equation}
A_m = \left(\frac{dQ'}{d\xi_m}\right)_P = T \left(\frac{d_iS}{d\xi_m}\right)_P = \mu_{r,m} - \mu_{p,m} = - ne\, \eta_m   \label{eq:Adef}
\end{equation}
where we also relate affinity to activation overpotential of a reduction reaction~\cite{bazant2013}.

The affinity can be viewed as a thermodynamic force,
$F_m = \frac{A_m}{T}$,
whose conjugate thermodynamic flux, $J_m=R_m$, is  the reaction rate 
\begin{equation}
R_m = \frac{1}{V} \frac{d\xi_m}{dt} = - \sum_i s_{r,m,i}\frac{dc_i}{dt} = \sum_j s_{p,m,j}\frac{dc_j}{dt} 
\end{equation}
(In thermodynamics~\cite{prigogine_book,prigogine1967introduction,glansdorff1971structure,kondepudi_book}, this is ``reaction velocity", $v_m$, but we adopt our previous notation for ``reaction rate"~\cite{bazant2013}, $R_m$, which also avoids any confusion with fluid velocity in liquid systems!) 
For thermodynamic consistency, the reaction rate must satisfy only two fundamental constraints:  
\begin{enumerate}
\item Equilibrium must correspond to detailed balance of the forward and backward rates 
\begin{equation}
A_m=0 \ \Leftrightarrow \ R_m = 0.  \label{eq:eq}
\end{equation}
\item Out of equilibrium, the net reaction must proceed in the direction of the affinity, which De Donder wrote expressed as positive irreversible entropy production per volume~\cite{prigogine1980nonequilibrium},
\begin{equation}
\sigma_m = A_m R_m = - \frac{ \eta_m I_m}{V}  > 0.  \label{eq:power}
\end{equation}
For Faradaic reactions, the integral reaction resistance must be positive, $\mathcal{R}_i=-\eta_m/I_m>0$, although the differential resistance, $\mathcal{R}_d=-\frac{d\eta_m}{dI_m}$, may have either sign, as discussed below.
\end{enumerate}
Prigogine~\cite{prigogine1947etude,prigogine_book} showed that closed reaction network is stable if the affinities decrease with each reaction extent, 
\begin{equation}
\mbox{ Stable:} \ \ \left(\frac{\delta{A}_m}{\delta\xi_n}\right)_P = T \left(\frac{\delta_i^2 S}{\delta\xi_m \delta\xi_n}\right)_P  < 0. 
\label{eq:prigstab}
\end{equation}
or equivalently that the irreversible entropy reaches a maximum in equilibrium, which follows from Gibbs' maximum entropy principle, Eq. (\ref{eq:genstab}), and De Donder's definition of affinity, Eq. (\ref{eq:Adef}). 

\subsection{ Linear Irreversible Thermodynamics of Reactions }

For a closed system in equilibrium, the irreversible entropy production vanishes. Close to equilibrium where LIT applies, Prigogine~\cite{prigogine1947etude} showed that the entropy production rate $P_e$ decreases and reaches a local minimum for any stationary non-equilibrium state\cite{prigogine_book,nicolis1977self,kondepudi_book},
\begin{equation}
P_e = \frac{d_iS}{dt} = \int_V \left(\sum_\alpha F_\alpha J_\alpha\right) dV > 0, \ \mbox{ Stable:} \ \frac{dP_e}{dt}<0    \label{eq:prigPstab}
\end{equation}
where the sum is over all pairs of conjugate forces $F_\alpha$ and fluxes $J_\alpha$, including each affinity and reaction rate.   The entropy production rate acts as a Lyapunov functional ($P_e>0$, $\dot{P}_e<0$), which can also determine the stability of non-equilibrium states~\cite{nicolis1977self,kondepudi_book}.

The analog of LIT fluxes for chemical reactions is the assumption of linear kinetics, which we express variationally as
\begin{equation}
R_m = k_m A_m = k_m \sum_i s_{m,i} \frac{\delta G}{\delta c_i}
\label{eq:linkin}
\end{equation}
where $k_m>0$ is a constant and $s_{m,j}= s_{p,m,j}  - s_{r,m,j}$ (positive stoichiometric coefficients for products, negative for reactants).  Although widely used, linear kinetics are strictly only valid near equilibrium in dilute mixtures~~\cite{prigogine1948affinity,bazant2013}.  Mass conservation equations take the form,
\begin{equation}
\frac{\partial c_i}{\partial t} = \sum_j  \left( \sum_{m} k_m s_{m,i}  s_{m,j}\right) \frac{\delta G}{\delta c_j}   \label{eq:multilinkin}
\end{equation} 
for a closed chemical reaction network. 

In this work, we focus instead on chemical reactions in open systems.
The standard phase-field model for a driven open system is the Allen-Cahn equation~\cite{allen1972ground,kom}, 
\begin{equation}
\frac{\partial c}{\partial t} = k_{res} A_{res} =  k_{res} \left(\mu_{res} - \frac{\delta G}{\delta c}\right)
 \label{eq:ACeq}
\end{equation}
where $\mu_{res}$ is the chemical potential of an external reservoir of species $c$.  The Allen-Cahn equation is usually applied to  non-conserved order parameters, such as the degree of solid-like order in liquid solidification, but when applied to chemical reactions, it corresponds to linear kinetics for a driven reaction.  As a result of this assumption, we shall see that the Allen-Cahn equation predicts the same spinodal region for a driven open systems as for closed equilibrium systems, Eq. (\ref{eq:gstab}), as shown in Fig. \ref{fig:I0}(a).  This is true even when diffusion is included in a combined Cahn-Hilliard/Allen-Cahn model~\cite{lamorgese2016spinodal}. As recognized by Prigogine, {\it nonlinear thermodynamics} are required for any departures from the equilibrium ``thermodynamic branch" of stability~\cite{prigogine1967introduction,glansdorff1971structure,prigogine1978coupling,kondepudi_book}, and we shall see that this also holds true for the stability of driven open systems, such as electrochemical cells.

\subsection{ Nonlinear Irreversible Thermodynamics of Reactions }

Huberman~\cite{huberman1976striations} added mass-action kinetics to the Cahn-Hilliard equation as a model for spinodal decomposition and pattern formation in a  reactive mixture,
\begin{equation}
\frac{\partial c}{\partial t} = \nabla \cdot L \nabla \frac{\delta G}{\delta c}   + R(c).  \label{eq:Huberman}
\end{equation}
Similar Ginzburg-Landau-type reaction-diffusion equations have been studied extensively in chemical physics as generic models of self-organization~\cite{desai_book}.  Glotzer {\it et al.} performed simulations and linear stability analysis of Eq. (\ref{eq:Huberman}) and reached the tantalizing conclusion that reactions could be used to alter the spinodal region and control pattern formation.  However, Lefever {\it et al.}~\cite{lefever1995comment} pointed out that the model is not thermodynamically consistent, since equilibrium ($\mu=$constant) is neither stationary ($\frac{\partial c}{\partial t} =0$) nor in detailed balance ($f=0$), and equilibrium states depend on the mobility or diffusivity.  Instead, the reaction rate must satisfy the two constraints given above, Eqs. (\ref{eq:eq})-(\ref{eq:power}), and the further assumption of linear kinetics (\ref{eq:linkin}) eliminates any effect on the spinodal region.  

A thermodynamically consistent linear stability analysis for general chemical reaction networks was performed by Carati and Lefever~\cite{carati1997chemical}, based on multi-component Cahn-Hilliard diffusion (\ref{eq:CHeq}) and (\ref{eq:mu1}) and a nonlinear reaction model converting species $i$ into species $j$:
 \begin{equation}
R_{ij} = f_r(\tilde{\mu}_i) - f_r(\tilde{\mu}_j), \ \ \  f_r^\prime > 0   \label{eq:RCL} 
\end{equation}
which upholds Eqs. (\ref{eq:eq})-(\ref{eq:power}).  Notably, they also considered open reaction networks with chemostats and predicted the possibility of ``chemical freezing" of phase separation, by two or more collectively autocatalytic reactions. 

Hildebrand, Mikhailov and Ertl~\cite{hildebrand1998nonequilibrium} analyzed general stochastic models of surface adsorption and also concluded that  ``thermal
adsorption and desorption processes do not prevent macroscopic
phase separation", but ``if, on the other hand, an energetically
activated process (such as photo-desorption) is
present, kinetic freezing of phase separation, leading to the
formation of stationary nonequilibrium structures, can occur," consistent with experiments and simulations on reactive monolayers~\cite{mikhailov2009nonequilibrium}.  In other words, the reaction must be driven, supplying external work.  Here, we focus on the possibility of using Faradaic reactions as the driving process. 

\subsection{ Variational Electrochemical Kinetics }

We shall modify some of these conclusions using more general models, based on transition-state theory for concentrated solutions and electrochemical systems~\cite{bazant2013}.  The theory is based on variational definitions of activity, $a_i=\gamma_i \tilde{c}_i$, activity coefficient $\gamma_i$, and excess chemical potential, $\mu_i^{ex}=k_BT\ln\gamma_i$:
\begin{equation}
\mu_i = \frac{\delta G}{\delta c_i} = \mu_i^\Theta + k_BT \ln a_i = k_BT \ln \tilde{c} + \mu_i^{ex}
\end{equation}
For the reaction, $M_i \to M_j$, the generalized Eyring rate is given by
\begin{equation}
R_{ij} = k_0 \left( e^{-(\mu_\ddag^{ex} - \mu_i)/k_BT} - e^{-(\mu_\ddag^{ex} - \mu_j)/k_BT}\right) 
=
\frac{k_0(K^\Theta_{ij} a_i  -  a_j)}{\gamma_\ddag}    \label{eq:genreact}
\end{equation}
where $K^{\Theta}_{ij}$ is the equilibrium constant and $\gamma_\ddag$ is the activity coefficient of the transition state, which generally depends on concentration, e.g. $\gamma_\ddag^{-1} = (1-\sum_l \tilde{c}_l)^s$ for $s$ excluded sites on a lattice. 

As a result, the model is more general than Eq. (\ref{eq:RCL}) and allows for negative differential resistance ($\frac{\partial R_{ij}}{\partial \mu_i} < 0$), as in Marcus kinetics, and solo-autocatalysis ($\frac{\partial R_{ij}}{\partial c_i} \neq 0$). The latter includes the important case of Butler-Volmer kinetics~\cite{bazant2013}: 
\begin{eqnarray}
I &=& neR = I_0 \left( e^{-\alpha \tilde{\eta}} - e^{(1-\alpha) \tilde{\eta}} \right),   \ \ \tilde{\eta}=\frac{ne\eta}{k_BT} \\
I_0 &=& \frac{nek_0(a_Oa_e)^{1-\alpha}a_R^\alpha}{\gamma_\ddag}   \label{eq:BV}
\end{eqnarray}
for the reduction reaction, $\mbox{O} + ne^- \to \mbox{R}$.

Once the reaction model is specified, the thermodynamically consistent set of reaction-diffusion equations takes the form~\cite{bazant2013},
\begin{equation}
\frac{\partial c_i}{\partial t} = \nabla\cdot \sum_j L_{ij} \nabla \frac{\delta G}{\delta c_j} 
+ \sum_m s_{i,m} R_m\left(\{ c_i \}, \left\{ \frac{\delta G}{\delta c_i}  \right\}\right)   \label{eq:reactdiff}
\end{equation}
where we write the $m$th reaction as $\emptyset \to \sum_i s_{i,m} \mbox{M}_i$.
\change{ This is the most general mathematical framework for concentration evolution, based on LIT fluxes and nonlinear irreversible thermodynamics for chemical reactions.}

\subsection{ Glansdorff-Prigogine Nonequilibrium Stability Theory }

Glansdorff and Prigogine derived a general linear stability condition for stationary non-equilibrium states of reactive mixtures far from equilibrium~\cite{glansdorff1970non,glansdorff1971structure,kondepudi_book}, based on variations of irreversible entropy production~\cite{glansdorff1954proprietes,prigogine1967introduction}.  They argued that the second variation of the entropy acts as a Lyapunov functional, $\mathcal{L}=-\frac{1}{2}\delta^2S$, which measures the ``distance" from a stationary state, $\mathcal{L}>0$, and thus decreases with time if it is stable, $\frac{d\mathcal{L}}{dt} < 0$. The stability criterion can be expressed as a constraint of positive excess entropy production~\cite{glansdorff1970non,glansdorff1971structure},
\begin{equation}
\mbox{ Stable:}\ \ \frac{d}{dt} \frac{\delta^2 S}{2} = \int_V  \left( \sum_\alpha \delta F_\alpha \cdot \delta J_\alpha \right) dV > 0,   \label{eq:glanstab}
\end{equation}
for an arbitrary set of conjugate forces $F_\alpha$ and fluxes $J_\alpha$. Besides reactions, there may also be contributions to excess entropy production from diffusion, electromigration, elastic deformation, heat conduction, etc.  The same result holds for any boundary conditions in which either the forces or fluxes are held fixed, causing the second variation of the entropy flow to vanish on the boundary. 

\begin{figure}[tb]
\begin{center}
\includegraphics[width=1.7in]{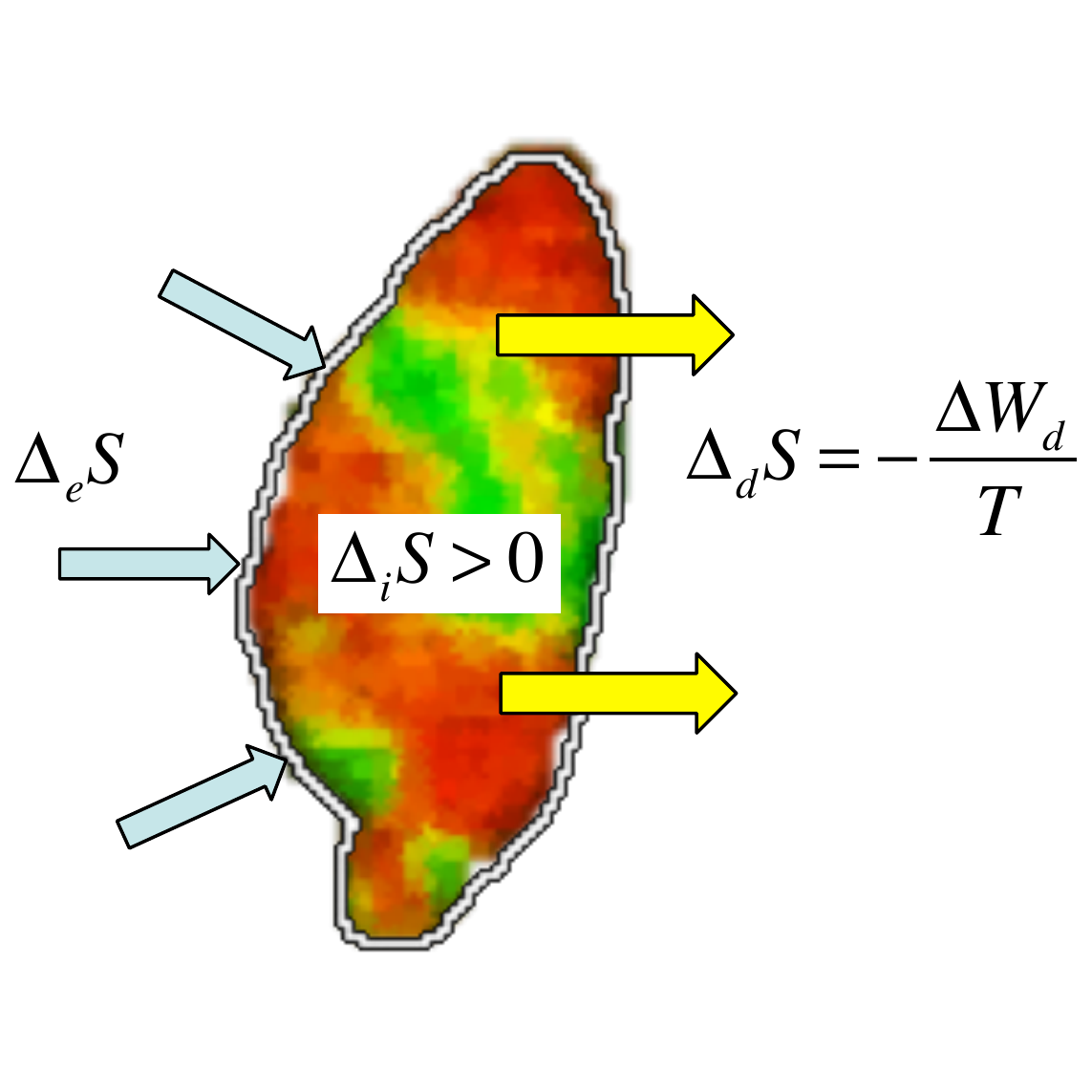}
\vspace{-0.3in}
\caption{ Three contributions to entropy production in a driven open system: (1) bulk irreversible entropy production, $\Delta_i S$, (2) entropy flow due to mass and energy flow through the boundary, $\Delta_e S$, and (3) driven entropy production, $\Delta_d S$, due to the work, $\Delta W_d$, done on the system by exchanging mass and energy directly between the external reservoirs and the interior bulk. The image shows a two-phase lithium iron phosphate nanoparticle driven far from equilibrium by an applied Faradaic current~\cite{lim2016origin} from Fig. ~\protect\ref{fig:lim} below.  \label{fig:entropy}  }
\end{center}
\end{figure}

\change{
Although the Glansdorff-Prigogine criterion (\ref{eq:glanstab}) follows from thermodynamically consistent mass and energy balances~\cite{kondepudi_book},  Keizer and Fox first expressed ``qualms" about its validity ~\cite{keizer1974qualms} and triggered a long  debate~\cite{glansdorff1974thermodynamic,fox1979irreversible,nicolis1979irreversible,fox1980excess}.   They provided counter-examples of auto-catalytic reaction networks in dilute solutions~\cite{keizer1974qualms,fox1980excess}, whose non-equilibrium steady states violate Eq. (\ref{eq:glanstab}), and yet could be described by Keizer's stochastic thermodynamics~\cite{keizer1976fluctuations,keizer1978thermodynamics,keizer1985heat,keizer2012statistical}. Glansdorff, Nicolis and Prigogine responded that different Lyapunov functions are possible depending on the choice of conservation laws~\cite{glansdorff1974thermodynamic,nicolis1979irreversible}, and pointed to Schl\"ogl's earlier derivation of Eq. (\ref{eq:glanstab}) based on  similar stochastic principles~\cite{schlogl1967statistical}, rooted in fluctuation-dissipation theorems for nonequilibrium states~\cite{lax1960fluctuations}.
}

\change{
We shall see that the problem has to do with driven, open systems.  In the counter-examples, nonequilbrium stationary states are constructed by fixing certain concentrations or production rates throughout the domain, but such ``chemostats" are neglected in the Glansdorff-Prigogine derivation~\cite{kondepudi_book}, which assumes an unconstrained system of reaction-diffusion conservation laws. The stability criterion cannot be expressed in terms of affinities by summing over all reactions, if any concentrations or rates are externally controlled.   } 

As shown in Fig. ~\ref{fig:entropy}, the theory generally does not account for {\it bulk} entropy flow \change{ from } distributed work done by ``active matter" or \change{by} the \change{direct} exchange of mass and energy with external reservoirs.  
\change{ Moreover, the traditional focus on reaction networks in dilute solutions obscures the rich new physics of driven reactions coupled with phase transformations. Here, we generalize the theory for concentrated solutions and show how driven reactions can control thermodynamic stability. }

\section{ Theory }
\label{sec:theory}

\subsection{ Wisdom from Stochastic Thermodynamics  }

Any theory of nonequilibrium thermodynamics for concentrated systems should be consistent with stochastic thermodynamics for the ideal limit of a dilute system with chemical reaction networks obeying mass action kinetics~\cite{sekimoto_book,keizer2012statistical,seifert2012stochastic,gaspard2004fluctuation,schmiedl2007stochastic,rao2016nonequilibrium}.  Rao and Esposito recently summarized this ``wisdom" and rigorously defined various forms of the {\it non-equilibrium Gibbs free energy} for open systems with driven chemical reactions~\cite{rao2016nonequilibrium},
\begin{equation}
\mathcal{G} = \mathcal{G}_{eq} + k_BT \mathcal{L}
\end{equation}
where $\mathcal{G}_{eq}$ is the local equilibrium free energy of a state that would reached if the external driving were stopped and the system were allowed to relax under the imposed constraints and $k_B\mathcal{L}$ is the ``relative entropy" between the equilibrium and nonequilibrium states. The relative entropy, also known as  the Kullback-Leibler divergence in information theory~\cite{kullback1951information}, is a non-negative measure of the ``information gain" between two probability distributions, which acts as a Lyapunov functional for the relaxation to local equilibrium.

The change in non-equilibrium free energy between two states,
\begin{equation}
\Delta\mathcal{G} = \Delta{W}_d - T \Delta_i S
\end{equation}
has contributions from external work and internal entropy production of opposite sign.  The work can be broken into irreversible and reversible parts, $\Delta W_d = \Delta_i W_d +  \Delta\mathcal{G}_{eq}$, where the reversible chemical work is equal to the change in local equilibrium free energy, as a result of exchanging bulk particles with the reservoirs. Combining these equations, we arrive at the central result of Rao and Esposito for irreversible chemical work in dilute mixtures~\cite{rao2016nonequilibrium},\begin{equation}
\Delta_i W_d = \Delta W_d - \Delta\mathcal{G}_{eq} = k_BT \Delta \mathcal{L} + T \Delta_iS
\end{equation}
The Second Law ($\Delta_i S > 0$) then implies a ``non-equilibrium Landauer principle"~\cite{rao2016nonequilibrium},
$\Delta_i W_d \geq k_B T \Delta\mathcal{L}$, which provides a lower bound on the irreversible external work associated with the fluctuation (the thermodynamic cost of information gain~\cite{takara2010generalization,esposito2011second}) that vanishes for transitions between equilibrium states ($\Delta \mathcal{L}=0$).  

\subsection{ Thermodynamic Stability of Driven, Open Systems }

Let us apply these principles more broadly to concentrated systems experiencing arbitrary forms of external driving work.  Enthalpy from inter-particle forces now  leads to nonlinear chemical diffusion and influences the enthalpies of reactions, both internal and external. As a result, the equilibrium free energy may lose convexity and lead to spinodal decomposition.  Thermodynamic stability will then be influenced by the driving work $\Delta W_d$ done on the bulk system, which may include contributions from heat transfer (e.g. radiation), mass transfer (e.g. chemical reactions with reservoirs), external forces (e.g. magnetic fields or mechanical work), or internal energy sources (e.g. swimming particles or other active matter).  

These contributions are neglected in the prevailing theory of nonequliibrium thermodynamics~\cite{kondepudi_book}.  Prigogine and collaborators described the thermodynamics of closed, internal reaction networks in what could be termed ``partially open" systems, in which entropy or energy exchange with external reservoirs occurs only through the boundaries.  In contrast, we consider ``fully open" driven systems, in which entropy flow and external work can also be distributed across the bulk system.

The key theoretical concept is the nonequilibrium free energy, $\mathcal{G}$. 
In some cases, it may be possible to construct $\mathcal{G}$ as a local state function in space and time, which depends on traditional intrinsic variables, such as chemical concentration, density, pressure and temperature, as well as intrinsic external driving forces or fluxes.  We have already discussed examples from the stochastic thermodynamics of chemical reaction networks~\cite{rao2016nonequilibrium,sekimoto_book}.   Nonequilibrium free energies have also been constructed for active suspensions of swimming particles~\cite{takatori2014swim,speck2014effective,takatori2015towards,takatori2016forces} and recently connected with stochastic thermodynamics~\cite{speck2016stochastic}.   For driven electrochemical systems, we have already constructed  $\mathcal{G}(c,I)$ for ion adsorption in a phase-separating electrode at constant concentration $c$ and constant current $I$ in Eqs. (\ref{eq:dg1})-(\ref{eq:wd1}).  Below, we shall explicitly construct the nonequilibrium free energy (via its variational derivatives) for a general homogeneous driven, open system.

In most cases, it is not possible to express $\mathcal{G}$ as a simple state function due to various  non-local, nonlinear processes in space and time, but we can still {\it define} the first variation of $\mathcal{G}$,  the response to an arbitrary fluctuation, as
\begin{equation}
\delta \mathcal{G} \equiv \delta W_d - T \delta_i S = \delta \mathcal{G}_{eq} + \delta_i W_d - T \delta_i S
\end{equation}
which can be integrated in time to obtain at least a path-dependent free energy, $\mathcal{G}(t)$. We can then identify a nonequilibrium steady state via  
\begin{equation}
\mbox{ Steady State:} \ \ \ \delta \mathcal{G}=0 \ \ \Rightarrow \ \ \delta_i S = \frac{\delta W_d}{T}
\end{equation}
which extends Gibbs' condition of thermal  equilibrium, $\delta_i S=0$, to account for driving work. The canonical example is a driven reaction network in detailed balance.    We can also write the steady state condition as $\delta S_{tot}=0$, where $S_{tot} = S_i + S_d$, where we define the change in {\it driving entropy}
\begin{equation}
\delta S_d = -\frac{\delta W_d}{T}
\end{equation}
associated with the external reservoirs, 
which has the opposite sign of the driving work, since work creates order and lowers entropy. 

In the thermodynamic limit of a continuous system, local fluxes and reactions maintain each infinitesimal bulk volume in quasi-equilibrium, leading to our first principle: 
\begin{equation}
\mbox{Local equilibrium:} \ \  \ \delta^2 \mathcal{G} = 
\delta^2 W_d - T \delta^2_i S > 0 
\end{equation}
which generalizes Gibbs' maximum entropy condition, Eq. (\ref{eq:genstab}), to account for driving work.  The local equilibrium condition can then be viewed as the Gibbs' criterion for the total entropy, $\delta^2 S_{tot} < 0$.

The definite sign of $\delta^2 \mathcal{G}$ allows the {\it second variation of nonequilibrium free energy to serve as a Lyapunov functional}, which implies thermodynamic stability if it decreases toward steady state ($\delta \mathcal{G}=0$) in response to fluctuations,
\begin{equation}
\mbox{ Stable:} \ \ \ \frac{d}{dt} \delta^2 \mathcal{G} =
\frac{d}{dt}\delta^2 W_d - T \frac{d}{dt}\delta^2_i S < 0    \label{eq:modglan}
\end{equation}
or $\frac{d}{dt}\delta^2 S_{tot} > 0$, which generalizes the Glandorff-Prigogine criterion of positive excess entropy production, Eq. (\ref{eq:glanstab}), to account for excess driving entropy production, $\frac{d}{dt}\delta^2 S_d$, or excess driving power, $\frac{d}{dt}\delta^2 W_d$.  Near equilibrium, this also generalizes Prigogine's principle of minimum entropy production, Eq. (\ref{eq:prigPstab}). 

The general stability criterion, Eq. (\ref{eq:modglan}), states that the excess entropy production from internal irreversible processes must exceed the excess driving power. Each term can take either sign. If the excess driving power is negative, it is possible to stabilize an ordered ``dissipative structure" having negative excess entropy production~\cite{nicolis1977self,prigogine1980nonequilibrium,kondepudi_book}. Conversely, an unstable system can be destabilized``chemically frozen"  in a disordered state by positive excess driving power.  These surprising phenomena appear to contradict the Duhem-Jougeut Theorem, which asserts that \change{a} system that is stable to diffusion is also stable to chemical reactions~\cite{prigogine_book,kondepudi_book}, but that is only true in a partially open system without bulk driving work\change{, under conditions derived below}. Different behavior is possible in fully open, driven systems.

\subsection{ \change{Variational} Linear Stability Analysis  }

In order to illustrate \change{these principles}, we now perform linear stability analysis on the most general thermodynamically consistent system of isothermal reaction-diffusion equations, Eq. (\ref{eq:reactdiff}), \change{ using the calculus of variations}. For any concentration fluctuations $\{ \delta c_i \}$ around \change{a} non-equilibrium base state, the simplest Lyapunov function is the L$^2$-norm of the perturbation,
\begin{equation}
\mathcal{L}_c = \frac{1}{2}\sum_i \int_V (\delta c_i)^2\, dV \geq 0
\end{equation}
which must decrease for a stable base state,
\begin{equation}
\mbox{ Stable:} \ \  \frac{d\mathcal{L}_c}{dt} 
= \sum_i \int_V  \left(\nabla\delta c_i \cdot \delta J_i + \sum_n s_{i,n} \delta c_i \delta R_n\right) dV < 0   
\label{eq:Lc} 
\end{equation}
where we use the divergence theorem and assume $\delta c_i=0$ on the boundary. 
\change{ The general linear stability result (\ref{eq:Lc}) resembles the Glansdorff-Prigogine criterion (\ref{eq:glanstab}) since it contains products of excess thermodynamic fluxes ($\delta J_i$, $\delta R_n$) and certain excess forces, but the latter are expressed in terms of concentration fluctuations ($\nabla \delta c_i$, $\delta c_i$), rather than fluctuations in proper thermodynamic forces  ($\delta \nabla \mu_i$, $\delta A_n$), which can only be derived for mass and energy balances in partially open systems~\cite{kondepudi_book}.  }

Assuming LIT fluxes and nonlinear reactions, Eq. (\ref{eq:reactdiff}), we can express the stability criterion as 
\begin{eqnarray}
\frac{d\mathcal{L}_c}{dt} &=& \sum_{ij} \int_V \left[ 
-\nabla\delta c_i\cdot \sum_l  \delta c_l 
\left( \frac{\partial L_{ij}}{\partial c_l}  \nabla{\mu}_j  
+ L_{ij} \nabla \frac{\delta \mu_j}{\delta c_l}\right)
\right. \nonumber \\
& & \left. +  \delta c_i \mathcal{A}_{ij} \delta c_j-(\nabla\delta c_i) \mathcal{D}_{ij} (\nabla\delta c_j)\right] dV  < 0    \label{eq:Lcstab}
\end{eqnarray}
where the first term involves fluctuations in the Onsager matrix,
\begin{equation}
L_{ij}=\frac{D_{ij} c_i}{k_BT}
\end{equation}
and only applies to inhomogeneous base states with $\nabla{\mu}_j\neq 0$.  The second term also vanishes for a homogeneous base state.  The remaining terms comprise a difference of two quadratic forms, 
whose physical meanings we now explain.

\subsection{ Autocatalytic Rate and Chemical Diffusion Tensors }

For slow diffusion, the stability of a homogeneous base state requires that the following tensor be negative definite:
\begin{equation}
\mathcal{A}_{ij} = \sum_n s_{i,n} \frac{\delta R_n}{\delta c_j} 
= \sum_n s_{i,n} \left( \frac{\partial R_n}{\partial c_j} + \sum_l \frac{\partial R_n}{\partial \mu_l} \frac{\delta \mu_l}{\delta c_j}\right)
< 0   \label{eq:Rauto}
\end{equation}
We refer to $\mathcal{A}$ as the  ``autocatalytic rate tensor", since it describes how reaction rates depend on the extents of both products ($s_{i,n}>0$) and reactants ($s_{i,n}<0$) within the system, excluding all reservoir species.  For linear stability with slow diffusion, a driven chemical reaction network must be auto-inhibitory, $\mathcal{A}<0$.
Prigogine's stability criterion based on affinities (\ref{eq:prigstab}) follows in the case of a closed system with linear kinetics (\ref{eq:linkin}), but Equation (\ref{eq:Rauto}) based on reaction rates is much more general.  

For slow reactions, a homogeneous base state is stable if the ``chemical diffusion tensor" is positive definite:
\begin{equation}
\mathcal{D}_{ij} = \sum_l L_{il} \frac{\delta \mu_l}{\delta c_j}   > 0    \label{eq:Dchemstab}
\end{equation}
Since the Onsager tensor, $L_{ij}$, and the tracer diffusion tensor, $D_{ij}$, are symmetric and positive definite, this implies that the Hessian tensor (\ref{eq:dG2}) must also be positive definite. Therefore, the requirement of positive definite chemical diffusion tensor, Eq. (\ref{eq:Dchemstab}), is equivalent to Gibbs' convexity criterion for the homogeneous free energy, Eq.  (\ref{eq:gstab}), which defines the classical chemical spinodal region for mixtures without external driving.  Alternatively, we can prove Onsager's reciprocal relations, $L_{ij} = L_{ji}$, as a consequence of Gibbs' maximum entropy principle, Eq. (\ref{eq:mustab}), and diffusional stability to concentration fluctuations, Eq. (\ref{eq:Lcstab}).  As usual in Thermodynamics,  axioms and theorems can often be interchanged, and which is more fundamental is in the eye of the beholder!

With these insights, the general stability criterion (\ref{eq:Lcstab}) clearly shows that {\it control of phase separation in a homogeneous mixture results from the competition between auto-catalysis and chemical diffusion}.  Outside the spinodal region ($\mathcal{D}>0$), a stable equilibrium system can undergo ``chemical melting" \change{(phase separation)}  if the reactions are sufficiently autocatalytic ($\mathcal{A}$ has large enough positive eigenvalue). Inside the spinodal region ($\mathcal{D}<0$), a unstable mixture can undergo "chemical freezing" \change{(stabilization}) if the reactions are sufficiently auto-inhibitory ($\mathcal{A}$ has large enough negative eigenvalue).  

\subsection{ Solo-autocatalysis and Differential Resistance }

The autocatalytic rate and chemical diffusion tensors can be further decomposed to clarify the connection with equilibrium thermodynamics:
\begin{eqnarray}
\mathcal{A} &=& S -\mathcal{R}^{-1} \, G^{\prime\prime}   \label{eq:AS} \\
\mathcal{D} &=& L \, G^{\prime\prime}
\end{eqnarray}
where we  define the ``solo-autocatalytic rate tensor",  
\begin{equation}
S_{ij} = \sum_n s_{i,n} \frac{\change{\partial} R_n}{\change{\partial} c_j},
\end{equation}
and the ``differential reaction resistance tensor",
\begin{equation}
\mathcal{R}^{-1}_{ij} = -\sum_n s_{i,n} \frac{\partial R_n}{\partial \mu_j}.
\end{equation}
If the reaction rates have no explicit concentration dependence ($S=0$) and positive differential resistances ($\mathcal{R} > 0$), then \change{ the Duhem-Jougeut Theorem holds: }linear stability ($\mathcal{A}<0$, $\mathcal{D}>0$) requires a convex equilibrium free energy ($G^{\prime\prime}>0$), and the chemical spinodal range remains unchanged by the driven reaction network. In particular, multicomponent, linear Allen-Cahn reaction kinetics (\ref{eq:multilinkin}) cannot alter the equilibrium spinodal region. Instead, the {\it control of phase separation} \change{ {\it by reactions} (in violation of the Duhem-Jougeut Theorem)} {\it requires either solo-autocatalaysis} ($S\neq 0$) {\it or negative differential resistance} ($\mathcal{R}<0$).

\subsection{ Nonequilibrium Gibbs Free Energy }

The preceding analysis allows us to variationally construct the nonequilibrium free energy that determines the stability of a uniform state.  In order to achieve stability in the long-wavelength limit, where reactions dominate diffusion ($\mbox{Da}>1$ defined below), the constraint of auto-inhibitory reactions, $\mathcal{A} < 0$, motivates the following definition, using Eq. (\ref{eq:AS}):
\begin{equation} 
\mathcal{G}^{\prime\prime} = G^{\prime\prime} - \mathcal{R} S   \label{eq:Gne}
\end{equation}
so that stability corresponds to $\mathcal{G}^{\prime\prime}>0$.  We see again that unless the reaction network is solo-autocatalytic, $S\neq0$, the equilibrium free energy will determine stability, since $\mathcal{G}^{\prime\prime}=G^{\prime\prime}$, and the reactions cannot alter the spinodal region. From Eq. (\ref{eq:Gne}), the second variation of $\mathcal{G}$ is determined by 
\begin{eqnarray} 
-\mathcal{R}^{-1}(\mathcal{G}^{\prime\prime}- G^{\prime\prime}) &=&   S  \\
\sum_{l,n} s_{i,n} \frac{\partial R_n}{\partial \mu_l} \frac{\delta^2(\mathcal{G}-G)}{\delta c_l \delta c_j}
&=& \sum_n s_{i,n} \frac{\change{\partial} R_n}{\change{\partial} c_j}   \label{eq:Gnoneq}
\end{eqnarray}
which can be used to determine stability.

In some special cases, Equation (\ref{eq:Gnoneq}) can be integrated to obtain the nonequilibrium free energy, or at least its first variational derivative, the nonequilibrium chemical potential, 
\begin{equation}
\mu^{noneq}_i = \frac{\delta \mathcal{G}}{\delta c_i}
\end{equation}
This is indeed possible for the simple Faradaic reaction model, $I(c,\mu)=I_0(c)(\tilde{\mu}_{res}-\tilde{\mu})$, for driven adsorption at constant current, considered above.  In that case, 
\begin{eqnarray}
\frac{\partial^2 \mathcal{G}}{\partial c^2} &=& \frac{\partial^2 g_h}{\partial c^2}  + \frac{\frac{\partial I}{\partial c}}{\frac{\partial I}{\partial \tilde{\mu}}} 
= \frac{\partial \mu_h}{\partial c}  + k_BT\frac{\partial}{\partial c} \frac{I}{I_0} \nonumber \\
\frac{\partial \mathcal{G}}{\partial c} &=&  \mu_h + (\mu_{res} - \mu_h) = \mu_{res}
\end{eqnarray}
we obtain the same nonequilibrium free energy as before, Eq. (\ref{eq:dg1}). \change{The} reservoir potential acts as the nonequilibrium chemical potential of the system, $\mu^{noneq} = \mu_{res}$, and the affinity of the reaction, $A = \mu_{res} - \mu_h$, 
 is equal to the difference between the nonequilibrium and equilibrium chemical potentials. 

\subsection{ Growth of Fourier Modes }

The variational analysis above holds for all infinitessimal fluctuations around a time-dependent base state.  Let us now consider the growth of sinusoidal perturbations, i.e. Fourier modes satisfying  $\nabla\delta c_i = \vec{k}_i\delta c_i$, which serve as a basis to represent arbitrary fluctuations. The Hessian tensor then takes the form
\begin{equation}
G^{\prime\prime}_{ij} = \bar{G}^{\prime\prime}_{ij} + \vec{k}_i\cdot K_{ij} \vec{k}_j
\end{equation}
and the Lyapunov functional grows as,
\begin{equation}
\frac{d\mathcal{L}_c}{dt} = \sum_{ij} \sigma_{ij} \int_V \delta c_i \delta c_j dV
\end{equation}
where 
\begin{eqnarray}
\sigma_{ij} &=& S_{ij} - \sum_l \left[ 
\mathcal{R}^{-1}_{il}\left( \bar{{G}}^{\prime\prime}_{lj} + \vec{k}_l\cdot K_{lj}\vec{k}_j\right) \right. \nonumber \\
& & \left. + \vec{k}_i\cdot L_{il}\bar{{G}}^{\prime\prime}_{lj} \vec{k}_j
+ (\vec{k}_i\cdot L_{il}\vec{k}_j)(\vec{k}_l\cdot K_{lj} \vec{k}_j) \right]    \label{eq:sigij}
\end{eqnarray}
is the growth rate matrix, 
\begin{equation}
\frac{\partial \delta c_i}{\partial t} = \sum_j \sigma_{ij} \delta c_j \ \ \Rightarrow \ \ 
\delta \vec{c} =  e^{\sigma t} \delta \vec{c}(t=0)
\end{equation}
which controls the exponential growth of collective fluctuations.

Equation (\ref{eq:sigij}) expresses the general principles above in yet another way.
Since $L, K > 0$,  regardless of equilibrium stability (signs of eigenvalues of $\bar{{G}}^{\prime\prime}$), the system is {\it destabilized by negative differential resistance} (negative eigenvalues of $\mathcal{R}^{-1}$) {\it or by solo-autocatalytic reactions} (positive eigenvalues of $S$), while it is {\it stabilized by solo-auto-inhibitory reactions} (negative eigenvalues of $S$).

\subsection{ Negative Differential Resistance }

The differential reaction resistances are usually assumed to be positive (like the integral resistance,  $\change{R}/A$), but this need not be the case in electrochemistry.  The most famous example is ``inverted region" of Marcus kinetics for outer-sphere electron transfer~\cite{marcus1956,marcus1993,bard_book,bazant2013}, where the differential resistance becomes negative at large over-potentials. The inverted region is a feature of bulk electron transfer reactions, although  integration over the Fermi distribution of electrons  restores positive differential resistance for Faradaic reactions at metallic electrodes~\cite{zeng2014MHC}.  

To the author's knowledge, this effect has never been considered in thermodynamic stability. From Eq. (\ref{eq:sigij}), we see that negative differential resistance acts like  backward diffusion with quadratic growth rate scaling as $-{\mathcal{R}}^{-1} k^2$ as $k \to \infty$, until it is cutoff by the quartic Cahn-Hilliard gradient penalty term. At long wavelengths $(k \to 0$),  it also changes the sign of the thermodynamic term $-{\mathcal{R}_p}^{-1} \bar{G}^{\prime\prime}$, which promotes stability inside and instability outside the equilibrium spinodal region.

\change{ 
\section{ Application to Driven Adsorption  }
\subsection{ Phase Field Model }
}

Returning to the physical picture in Sec. \ref{sec:phys}, let us consider the simplest case of driven, solo-autocatalytic adsorption described by a phase-field model~\cite{bazant2013},
\begin{equation}
\frac{\partial c}{\partial t} = \nabla \cdot L \nabla \mu + R(c,\mu,\mu_{res}), \ \ \ \mu = \frac{\delta G}{\delta c}   \label{eq:ACHR}
\end{equation}
with isotropic $L, K > 0$.  From Eq. (\ref{eq:sigij}), the growth rate of the $\vec{k}$ Fourier mode is 
\begin{equation}
\sigma = S - \left( \bar{G}^{\prime\prime} + K k^2 \right)\left( \mathcal{R}_p^{-1} + L k^2 \right)
\end{equation}
where the coefficients are all scalars:
\begin{equation}
S = \frac{\partial R}{\partial c}, \ \  \mathcal{R}_p^{-1} = - \frac{\partial R}{\partial \mu},
\ \ \bar{G}^{\prime\prime}= \frac{d\mu_h}{dc} = \frac{d^2g_h}{dc^2}.
\end{equation}
Let us analyze in detail the possibility of suppression of phase separation ($\sigma<0$) as the system is driven by the adsorption reaction through the spinodal region ($\bar{G}^{\prime\prime}<0$), in the typical case of positive differential resistance ($\mathcal{R}_p^{-1}>0$).  

The growth rate has a simple dimensionless form,
\begin{equation}
\tilde{\sigma} = \tilde{S} + (1-\tilde{k}^2)(\mbox{Da} + \tilde{k}^2)    \label{eq:sigt}
\end{equation}
where the wavenumber 
\begin{equation}
\tilde{k}^2 = \frac{ K k^2}{|\bar{G}^{\prime\prime}|} = (\ell k)^2
\end{equation}
is scaled to a characteristic length scale,  
\begin{equation}
\ell\change{^2} = \frac{K}{|\bar{G}^{\prime\prime}|} = \frac{\kappa}{c_s k_BT} \left| \frac{d\tilde{\mu}_h}{d\tilde{c}}\right|^{-1}
\end{equation}
which is proportional to the phase boundary thickness and diverges at the spinodal limits ($\tilde{c}=c/c_s$, $\tilde{\mu}=\mu/k_BT$).  The growth rate and solo-autocatalytic rate
\begin{eqnarray}
\tilde{\sigma} &=& \frac{ K\sigma}{L |\bar{G}^{\prime\prime}|^2} = \sigma \tau_d \\
\tilde{S} &=& \frac{ KS}{L |\bar{G}^{\prime\prime}|^2} = S \tau_d
\end{eqnarray}
are scaled to the characteristic time scale for {\it backward} diffusion,
\begin{equation}
\tau_d  = \frac{ K}{L |\bar{G}^{\prime\prime}|^2} = \frac{\ell^2}{|\bar{\mathcal{D}}|}
\end{equation}
where $\bar{\mathcal{D}} = L\bar{G}^{\prime\prime} < 0$ is the chemical diffusivity, which vanishes at the spinodal limits.

\begin{figure}
\begin{center}
\includegraphics[width=2.6in]{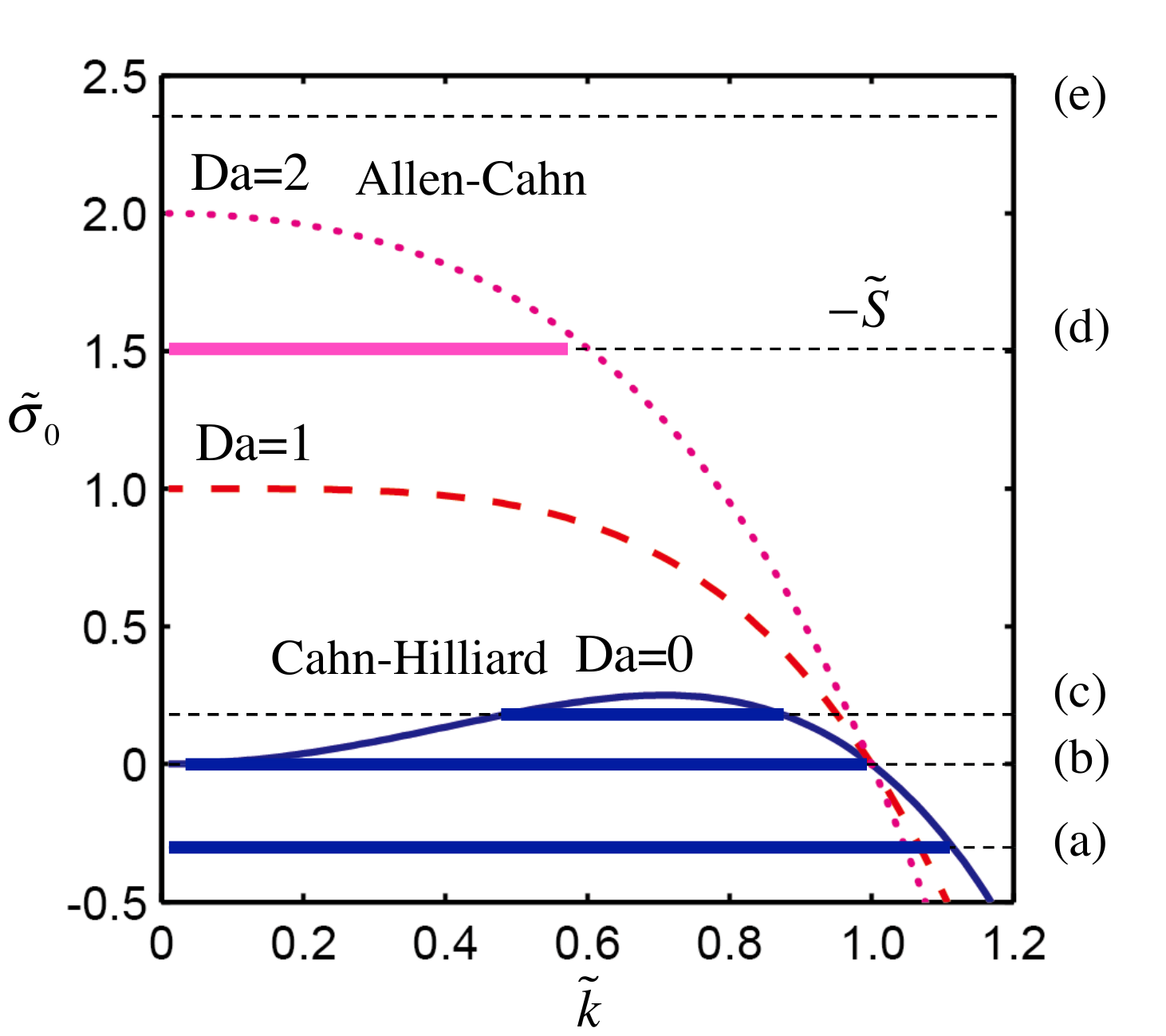}
\end{center}
\caption{ Control of phase separation by driven adsorption (or electro-autocatalysis). The band of unstable modes (orange) can be constructed graphically from the intersection of $-\tilde{S}=\tilde{\sigma}_0=(1-\tilde{k}^2)(\mbox{Da} + \tilde{k}^2)$. For a solo-autocatalytic reaction (a), $\tilde{S}>0$, the instability is Allen-Cahn-like for all $\mbox{Da}$. For a non-autocatalytic reaction (b), $\tilde{S}=0$, the instability  is Cahn-Hilliard-like for $\mbox{Da}\ll 1$ and Allen-Cahn-like for $\mbox{Da}\gg1$. For a weakly auto-inhibitory reaction (c) with fast diffusion, $\mbox{Da}\ll 1$, a narrow band of modes at finite wavelength can be selected. For strongly auto-inhibitory  reactions, (d) the instability is Allen-Cahn-like for fast reactions, $\mbox{Da}\gg 1$, or (e) supressed above a critical reaction rate.
\label{fig:spectrum} }
\end{figure}

As usual in chemical engineering, the relative importance of reactions compared to diffusion is measured by the Damk{\"o}hler number~\cite{deen_book,singh2008}, 
\begin{equation}
\mbox{Da} = \frac{K}{\mathcal{R}_p L |\bar{G}^{\prime\prime}|} = \frac{\tau_d}{\tau_r}    \label{eq:Dadef}
\end{equation}
which is the ratio of the diffusion and reaction time scales, only here diffusion is backward ($\mathcal{D}<0$)~\cite{lamorgese2016spinodal}, and the characteristic reaction time is  
\begin{equation}
\tau_r = \frac{ \mathcal{R}_p } { |\bar{G}^{\prime\prime}|}= \left| \frac{ d\mu_h}{dc} \frac{\partial \change{R}}{\partial \mu} \right|^{\change{-1}}
\end{equation}
which diverges at the spinodal limits (``critical slowing down"). 
For a non-autocatalytic reaction, $\tilde{S}=0$,  the Damk{\"o}hler number controls the shape of the growth-rate spectrum, $\tilde{\sigma}_0=(1-\tilde{k}^2)(\mbox{Da} + \tilde{k}^2)$, which interpolates between the Allen-Cahn-like fast-reaction limit,
$\sigma_0 \tau_r = \frac{\tilde{\sigma}_0}{\mbox{Da}} \sim 1- \tilde{k}^2$ for $\mbox{Da} \gg 1$,
and the Cahn-Hilliard-like fast-diffusion limit,
$\sigma_0 \tau_d = \tilde{\sigma}_0 \sim (1-\tilde{k}^2)\tilde{k}^2$ for  $\mbox{Da} \ll 1$. Due to critical slowing down of diffusion, the Allen-Cahn-like instability dominates near the spinodal limits ($\mbox{Da} \to 0$), while the Cahn-Hilliard-like instability may arise only deep into the spinodal region.  Such phenomena were recently studied by Lamorgese and Mauri~\cite{lamorgese2016spinodal} for a non-autocatalytic reaction with  linear Allen-Cahn kinetics~\cite{bazant2013}, in which case the spinodal limits of phase separation cannot be altered.

In contrast, control of phase separation is possible with nonlinear phase-field reaction kinetics~\cite{bazant2013}.  The most unstable wavenumber is generally given by
\begin{equation}
\tilde{k}_{max} = \sqrt{ \frac{ 1 - \mbox{min}\{\mbox{Da},1\}}{2}}
\end{equation}
The solo-autocatalytic rate shifts the growth-rate spectrum by a constant and selects the band of unstable modes via $\sigma_0(\tilde{k}) > -\tilde{S}$, as shown in Fig. ~\ref{fig:spectrum}.

\begin{figure*} 
\includegraphics[width=\linewidth]{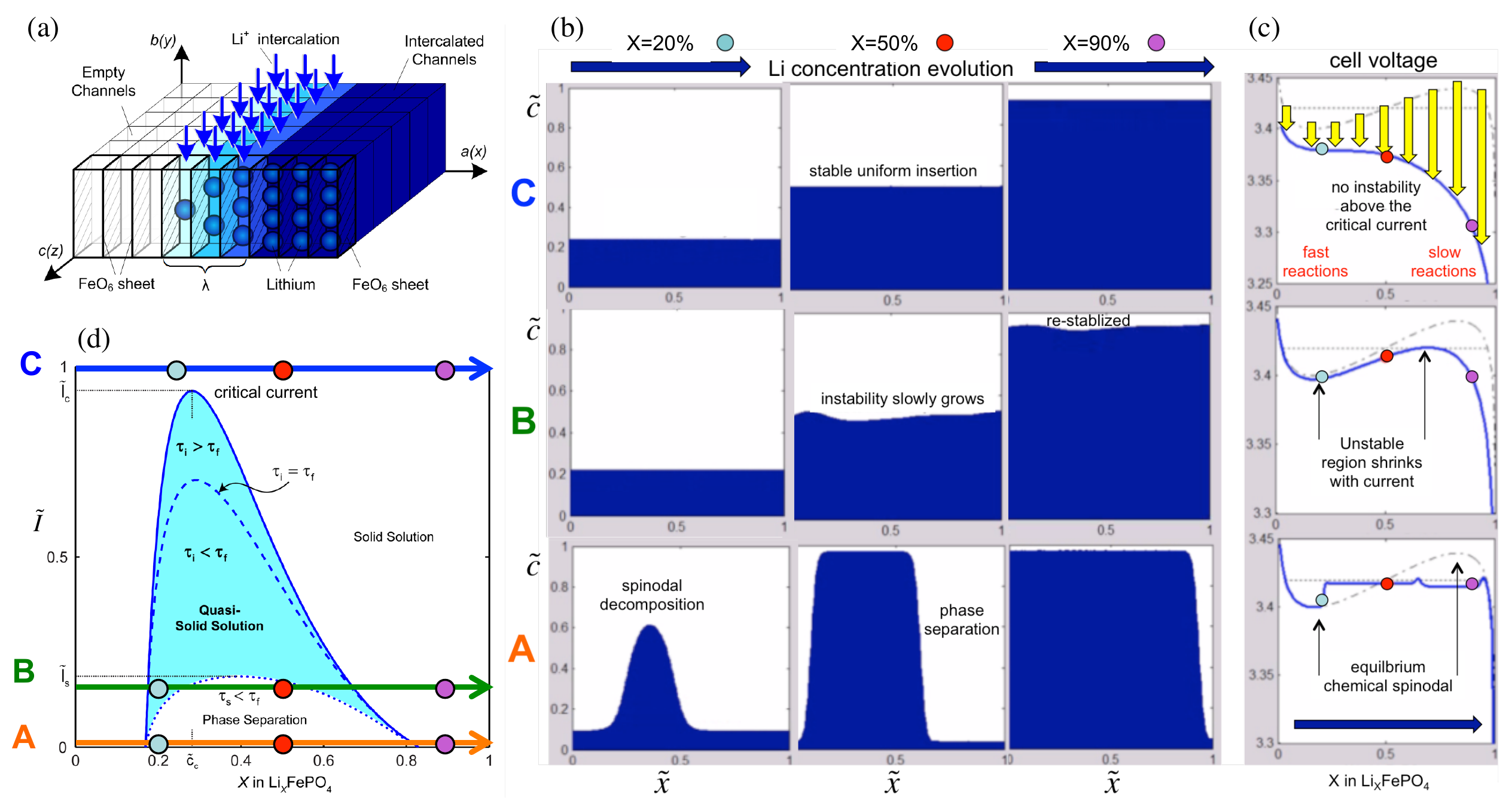}
\caption{ Electro-autocatalytic control of phase separation in an Allen-Cahn Reaction model for  lithium insertion in LFP~\cite{bai2011}, based on phase-field Butler-Volmer kinetics with regular solution thermodynamics~\cite{bazant2013} (neglecting coherency strain~\cite{cogswell2012}, $B=0$). (a) Sketch of the insertion reaction and depth-average concentration. (b) Simulations of the concentration profile and (c) cell voltage (versus a constant lithium reference) for three different currents, which are indicated as linear paths in the  (d) ``nonequilibrium phase diagram" of thermodynamic stability, $\tilde{I}>\tilde{I}_c(\tilde{c})$, in the plane of applied current $\tilde{I}=I/I_r$ and homogeneous concentration $X=\tilde{c}$.  [Adapted from ~\citet{bai2011}]  \label{fig:bai}}
 \end{figure*}

\subsection{ Critical Rate to Suppress Phase Separation }

The stability criterion, $\mbox{max}_{\tilde{k}}\tilde{\sigma} < 0$, can be expressed as a bound on the (negative) solo-autocatalytic rate,
\begin{equation}
S \tau_r = \frac{\tilde{S}}{\mbox{Da}} < - \frac{(1+\mbox{min}\{\mbox{Da},1\})^2}{4\, \mbox{min}\{\mbox{Da},1\} } = - F(\mbox{Da}) \leq 0
\end{equation}
or with dimensions restored,
\begin{equation}
\frac{\partial R}{\partial c} < - \left| \frac{ d\mu_h}{dc} \frac{\partial R}{\partial \mu} \right|\, F(\mbox{Da}) \leq  0  \label{eq:dvdcstab}
\end{equation} 
where $F=1$ for fast reactions ($\mbox{Da}\geq 1$) and $F \sim (4\mbox{Da})^{-1} \to \infty$ for slow reactions ($\mbox{Da}\ll 1$). In the latter limit, chemical diffusion promotes phase separation, so an increasingly negative (auto-inhibitory) solo-autocatalytic rate is required to maintain stability.
 
At constant reservoir potential, Equation (\ref{eq:dvdcstab}) implies an upper bound on the total autocatalytic rate,
\begin{equation}
\mathcal{A} = \left(\frac{dR}{dc}\right)_{\mu_{res}} = \frac{\partial R}{\partial c} + \frac{\partial R}{\partial\mu} \frac{d\mu_h}{dc}
<  - \left| \frac{ d\mu_h}{dc} \frac{\partial R}{\partial \mu} \right| \, 
\frac{(1-\mbox{min}\{\mbox{Da},1\})^2}{4\, \mbox{min}\{\mbox{Da},1\} } 
\end{equation}
For fast reactions, $\mbox{Da}>1$, we recover the constant-potential stability criterion, $\mathcal{A} < 0$, discussed in Section ~\ref{sec:phys}, Eq. (\ref{eq:Rstab}).

At constant current, the stability criterion can be expressed as a lower bound on the derivative of reservoir potential with \change{respect to} reaction extent (or time),
\begin{equation}
\left( \frac{\partial \mu_{res}}{\partial c}\right)_{\change{R,c}} > \frac{ \mathcal{R}_r }{\mathcal{R}_p} \left| \frac{ d\mu_h}{dc} \right| 
\,  \frac{(1-\mbox{min}\{\mbox{Da},1\})^2}{4\, \mbox{min}\{\mbox{Da},1\} }    \label{eq:muresstab}
\end{equation}
where we define the reactant differential reaction resistance,
\begin{equation}
\mathcal{R}_r^{-1} = \frac{\partial R}{\partial \mu_{res}}
\end{equation}
which is positive for driven adsorption on interfaces or electrodes.   The reactant and product differential resistances are equal for Carati-Lefever kinetics (\ref{eq:RCL}), which includes the limit of linear kinetics (\ref{eq:linkin}), but for our more general model (\ref{eq:genreact}), which includes Butler-Volmer kinetics~\cite{bazant2013}, they are typically different, $\mathcal{R}_p \neq \mathcal{R}_r$.
For fast reactions, $\mbox{Da}>1$, Equation (\ref{eq:muresstab}) reduces to the constant-current stability criterion, $\frac{\partial \change{R}}{\partial \mu_{res}} > 0$, discussed in Section ~\ref{sec:phys}, Eq. (\ref{eq:murstab}).

\subsection{ Control of Phase Separation by Electro-autocatalysis }

Finally, we are ready to apply our general stability theory to electro-autocatalysis. Consider generalized Butler-Volmer kinetics, Eq. (\ref{eq:BV}), for symmetric charge transfer ($\alpha=\frac{1}{2}$) as a model for cation reduction and adsorption or intercalation from an electrolyte reservoir to a cathode surface~\cite{bazant2013}:
\begin{equation}
I = neR = 2 I_0 \sinh\left(\frac{\tilde{\mu}_{res}-\tilde{\mu}}{2}\right)
\end{equation}
where the exchange current density 
\begin{equation}
I_0 = I_r(1-\tilde{c})e^{\tilde{\mu}/2}
\end{equation}
makes the reaction solo-auto-inhibitory ($S<0$)
and thus capable of suppressing phase separation, as a result of lattice crowding, $\gamma_\ddag=(1-\tilde{c})^{-1}$.  The prefactor, $I_r=ne k_0 \sqrt{a_{res}a_e}$, is constant, if we assume constant chemical activities of the electrolyte and electrons (fixed band structure). The effective reservoir chemical potential, $\mu_{res}$, is then controlled by the cathode potential, since the activation over-potential is $\eta = (\mu-\mu_{res})/ne$. Assuming that the anode is held at constant potential, the cell voltage is $V = V_0 - ne\mu_{res}$, where $V_0$ is the open circuit voltage when $\mu_{res}=\mu=0$.

Using  
\begin{equation}
\frac{\partial I}{\partial \tilde{c}} = - \frac{I}{1-\tilde{c}}, \ \ \  \ \ 
\frac{\partial I}{\partial \tilde{\mu}} = \frac{I}{2} - \sqrt{ I_0^2 + \left(\frac{I}{2}\right)^2}    \label{eq:dIdmu}
\end{equation}
the stability criterion, Eq. (\ref{eq:dvdcstab}), 
\change{ implies 
}
\begin{equation}
I \change{>} \frac{2 I_r (1-\tilde{c}) e^{\tilde{\mu}_h/2}}{\sqrt{ \left[ 1 + \change{2}\left((1-\tilde{c}) \left| \frac{d\tilde{\mu}_h}{d\tilde{c}}\right| F(\mbox{Da})\right)^{-1} \right]^2 -1}}  \label{eq:curstab}
\end{equation}
\change{
This is an implicit equation for the (positive) critical current  $I_c(\tilde{c})$ that suppresses phase separation, since the Damk\"ohler number  is current-dependent:
\begin{equation}
\mbox{Da} = \frac{K \left(\sqrt{ (2I_0)^2 + I^2}-I\right)}{2ne k_B T |\bar{\mathcal{D}}|}
\end{equation}
according to Eqs. (\ref{eq:Dadef}) and (\ref{eq:dIdmu}).
}

\change{
\subsection{ Role of Diffusion }
There are two different regimes of stability, depending on the importance of diffusion compared to reactions, as defined by the Damk\"ohler number:
\begin{itemize}
\item {\it Slow diffusion}. For relatively fast reactions ($\mbox{Da}\geq 1$), the critical current is given by the bound in Eq. (\ref{eq:curstab}) with $F=1$. Interestingly, the stability criterion is independent of the diffusivity for all $\mbox{Da}\geq 1$, not only in the asymptotic  limit of slow diffusion ($\mbox{Da}\gg 1$). 
\item {\it Fast diffusion.} Once the diffusivity surpasses a critical value defined by $\mbox{Da}>1$, the critical current increases.  Destabilizing chemical diffusion then begins to dominate over stabilizing electrocatalysis.
In the limit of fast diffusion ($\mbox{Da}\ll 1$, $F \sim (4\mbox{Da})^{-1}$), the critical current has the asymptotic form
\begin{equation}
I_c(\tilde{c}) \sim \left(\frac{nek_BT D}{4K}\right) \tilde{c}(1-\tilde{c})\left(\frac{d\tilde{\mu}_h}{d\tilde{c}}\right)^2 = \frac{nec_s(1-\tilde{c})}{4\tau_d}   \label{eq:Icdiff}
\end{equation}
which scales with the diffusion current, $nec_s/\tau_d$ (full capacity per diffusion time).
Although the critical current does not depend on rate-constant prefactor, $I_r$,  in this limit, it still depends on the concentration-dependence of the reaction rate (electro-autocatalysis).  Indeed, the general stability criterion (\ref{eq:dvdcstab}) for $\mbox{Da}\ll 1$ can still be expressed as a bound on solo-auto-inhibition
\begin{equation}
\frac{\partial R}{\partial c} < -\frac{ |\bar{\mathcal{D}}|}{4K} \left| \frac{d\mu_h}{dc}\right| 
\end{equation}
which takes the form of Eq. (\ref{eq:Icdiff}) using Eq. (\ref{eq:dIdmu}).
\end{itemize}
}

\change{
\subsection{  Regular Solution Thermodynamics}
}

The critical current, $I_c(\tilde{c},\mbox{Da})$, separates the stable and unstable regions of the ``non-equilibrium phase diagram" of current versus concentration \change{(and temperature)}, which are traversed during the dynamics. \change{An example is} shown in Fig. \ref{fig:bai}(d) for Butler-Volmer kinetics with regular solution thermodynamics~\cite{bai2011,bazant2013}:
\begin{eqnarray}
\tilde{\mu}_h&=&\ln \frac{\tilde{c}}{1-\tilde{c}} + \tilde{\Omega}(1-2\tilde{c})+\tilde{B}(\tilde{c}-X) \label{eq:regsol} \\  
\frac{d\tilde{\mu}_h}{d\tilde{c}} &=& \frac{1}{\tilde{c}(1-\tilde{c})} - 2\tilde{\Omega} + \tilde{B}
\end{eqnarray}
where $\Omega$ is the enthalpy of mixing particles and vacancies. \change{ For solid-state intercalation, t}he last term derives from the elastic coherency strain energy for small fluctuations~\cite{cogswell2012,cahn1962sd}, where $X$ is the average concentration, and $\tilde{c}=X$ for a homogeneous base state. \change{ Without strain ($\tilde{B}=0$), equilibrium in this model ($\mu=$constant) corresponds to the Frumkin isotherm for adsorption with lateral forces~\cite{bockris2000modern}.}

\begin{figure*}[t]
\begin{center}
\includegraphics[width=0.9\linewidth]{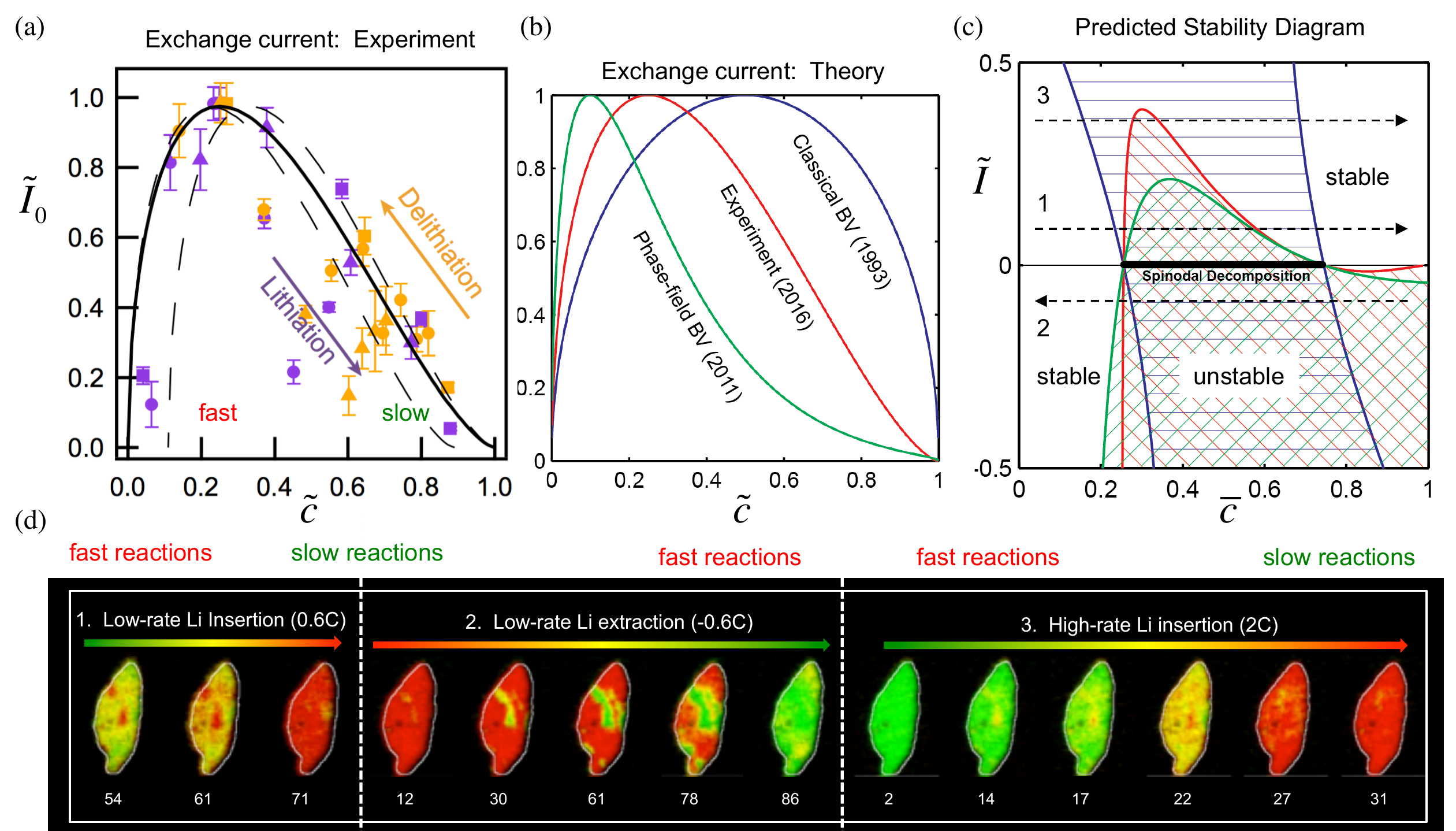}
\caption{ Direct experimental evidence for the control of phase separation by electro-autocatalysis in single nanoparticles of lithium iron phosphate, obtained by {\it in operando} scanning transmission x-ray microscopy (STXM)~\cite{lim2016origin}. (a)  Exchange current versus local surface concentration, obtained by pixel-level analysis of STXM movies of lithium evolution, (b) compared with the exchange current for Butler-Volmer kinetics from our original phase-field model~\cite{bai2011} and traditional porous electrode theory~\cite{doyle1993}.  (c) Predicted linear stability diagram versus current and composition for the exchange current curves in (b) from models and experiment. (d) Typical sequence of STXM images showing the lithium concentration profile ($X=0, 0.5, 1.0$ in Li$_X$FePO$_4$ for green, yellow, red) in a $\sim1\mu$m sized platelet particle (150nm thick in the depth direction) during cycling at different rates. At a moderate insertion rate (0.6C$=100$min to full capacity) some some phase separation occurs, which is enhanced significantly during extraction at the same rate (-0.6C). Next, at a high current (2C$=30$min discharge) above the critical insertion rate, phase separation is suppressed (``electrochemical freezing"), and uniform, stable insertion is observed.   [Adapted from ~\citet{lim2016origin}] \label{fig:lim}}
\end{center}
 \end{figure*}

For this reaction model, Figure \ref{fig:bai} shows simulations of the Allen-Cahn Reaction equation, Eq. (\ref{eq:ACHR}) with $L=0$,  which confirm the predictions of the stability theory~\cite{bai2011}. The concentration profiles in (b) develop long-wavelength fluctuations (set by the geometry) which grow as the system passes through the unstable region of nonequilibrium phase diagram (d), as signified by of increasing battery voltage in (c) ($\frac{dV}{dc} \sim - \frac{d\mu_{res}}{dc}>0$). The fluctuations decay as soon as the system re-enters a stable region of (d), and the voltage begins to decrease again in (c).

The control of phase separation is demonstrated by three currents in Fig. \ref{fig:bai}: (A) For small currents, $I \ll I_c$, the system overshoots the phase-separated equilibrium voltage plateau, undergoes spinodal decomposition, and then closely follows the voltage plateau, offset only by a small activation overpotential, associated with the moving phase boundary, or ``intercalation wave"~\cite{singh2008,bai2011}. (B) At larger currents, $I < I_c$, the instability is hindered, and the system behaves as a ``quasi-solid solution" in the unstable regions of increasing voltage. (C) Above the critical current, $I>I_c$, the homogeneous solid solution is stable, and the voltage decreases monotonically as a result of the concentration-dependent activation overpotential (yellow arrows in (c)).

\section{ Experimental Evidence }
\label{sec:expt}

\subsection{ Lithium Iron Phosphate Intercalation Kinetics }

Strong experimental support for the present theory has recently been achieved, after a decade of research in the field of Li-ion batteries.   Many battery materials exhibit multiple phases with varying composition, voltage, and temperature~\cite{whittingham2004,bruce2008,dunn2011}, and our nonlinear phase-field reaction model,  Eqs. (\ref{eq:genreact}) and (\ref{eq:reactdiff}), was first developed for this application, starting in 2007~\cite{bazant2013,singh2008}.  In the prototypical case of lithium iron phosphate (LFP), the model led to the surprising prediction that insertion reactions can suppress phase separation in nanoparticles above a critical current~\cite{bai2011}, even in the presence of heterogeneous nucleation~\cite{bai2011,cogswell2013} and elastic coherency strain ~\cite{cogswell2012}, although the underlying mechanism -- electro-autoinhibition -- was not explained until now.  

This theoretical prediction helped to explain the  dramatic reversal of fortune of LFP as a  battery material.
In the original paper on LFP, Goodenough and co-workers concluded that ``this material
is very good for low-power applications; at higher current
densities there is a reversible decrease in capacity that, we
suggest, is associated with the movement of a two-phase
interface"\cite{padhi1997}.   Indeed, phase separation is undesirable since it damages the crystal with coherency strain and lowers the rate capability by storing lithium in non-reactive stable phases~\cite{bai2011,bazant2013}.  Within a few years, however, LFP was reformulated as nanoparticles~\cite{bruce2008} with conductive coatings and demonstrated ultrafast ($<10$ sec) discharge without clear signs of phase separation in the voltage profile~\cite{kang2009}, despite the assumption of two-phase ``shrinking core" particles in prevailing mathematical models~\cite{srinivasan2004,dargaville2010}. 

The new theory was controversial, however, and competing hypotheses were made.  The existence of a ``solid solution pathway" of uniform insertion (and extraction) in LFP was suggested, on the basis that classical nucleation theory would prohibit nucleation and growth ~\cite{Malik2011}.   On the other hand, \change{our} phase-field model predicted that phase separation can nucleate at surfaces and collapsed experimental data for the size-dependent nucleation barrier~\cite{cogswell2013}.   Phase separation was later observed {\it in situ} in LFP porous electrodes~\cite{chueh2013}, and compared with phase-field porous electrode simulations~\cite{li2014current,ferguson2014}.

In 2014, three groups reported the first experimental evidence for the suppression of phase separation in LFP at high insertion rates~\cite{zhang2014rate,niu2014situ,liu2014}, although none could settle the question of the mechanism.  Zhang et al.~\cite{zhang2014rate} and Liu et al.~\cite{liu2014} used {\it in situ} synchrotron diffraction to measure the volume averaged Li$^+$ site occupation distribution (Fe$^{+3}$/Fe$^{+2}$ redox state). Each study found a transition from two-phase to solid-solution transformation above a critical current~\cite{zhang2014rate,liu2014} but could not observe the concentration profiles or reaction kinetics.   Meanwhile, Niu et al.~\cite{niu2014situ} were the first to directly observe nonequilibrium solutions in LFP nanowires, although the situation was artificial and could not shed light on the reason for their stabilization.

In 2016,  Lim et al.~\cite{lim2016origin} achieved a remarkable first test of the theory by {\it in operando} scanning transmission x-ray microscopy of single LFP nanoparticles in a microfluidic electrochemical cell.  The two-dimensional  lithium concentration evolution was directly observed with nanoscale resolution over the active facet of platelet-like nanoparticles, during realistic conditions of battery cycling.  The massive dataset of pixels from many movies of concentration evolution allowed the team to extract the local current density, and hence the exchange rate, versus local concentration, and the experimental curve (Fig. ~\ref{fig:lim}(a)) is asymmetric and similar to the original phase-field model of Butler-Volmer kinetics~\cite{bazant2013,bai2011,cogswell2012}, and different from the symmetric form, $I_0\sim\sqrt{\tilde{c}(1-\tilde{c})}$, assumed in traditional diffusion models (Fig. ~\ref{fig:lim}(b)).   The experimental and phase-field insertion reactions are solo-autoinhibitory across the spindoal region, which leads to suppression of phase separation above a critical insertion current and enhanced instability during extraction (Fig. ~\ref{fig:lim}(c)), as explained in Fig.~\ref{fig:spin}. In contrast, the symmetric reaction model predicts phase separation at all currents.   

As shown in Fig. ~\ref{fig:lim}(d), the data for repeated cycling of single nanoparticles confirm the theoretical prediction for the asymmetric exchange current. Lithium insertion at a moderate rate (0.6C$=$100 min. discharge) exhibits quasi-solid solution behavior with non-uniform concentration, while extraction at the same rate produces clear phase separation. Re-insertion in the same nanoparticle at a higher rate (2C$=$capacity in 30 min. discharge) leads to stable, uniform filling, but high-rate extraction (not shown) still leads to phase separation. When the current is turned off at intermediate concentrations (not shown), spinodal decomposition leads to striped equilibrium phase patterns, also predicted by the model with coherency strain~\cite{cogswell2012}.  Previous models~\cite{doyle1993,newman_book} could not predict any of these observations.

\subsection{ Lithium Peroxide Electrodeposition Kinetics }

Our general theory can also be applied to epitaxial surface growth or electrodeposition, where the surface height $h(x,t)$ acts as a surface concentration $c(x,t)$ integrated over the depth of the deposit~\cite{horstmann2013}.  In that case, the free energy functional $G[h]$ contains different physical effects, such as orientation-dependent surface energy and discrete stable monolayers, but the reaction kinetics can still be described by phase-field Butler-Volmer kinetics.  In this context, the instability of a uniformly growing film to ``phase separation" corresponds to the homogeneous nucleation and growth of islands, which can be controlled by electro-autocatalysis, according to the same principles revealed by studies of ion intercalation. 

In the case of lithium peroxide deposition in Li-air battery cathodes, the model successfully predicted a transition from island growth at low rates to homogeneous, random deposition at moderate rates, in good agreement with experimental voltage profiles and {\it ex situ} observed growth morphologies~\cite{horstmann2013}.  This is the surface-growth analog of suppression of phase separation in driven adsorption.  Although the local current density and exchange current could not be measured, this experiment shows the generality of the present theory, which is by no means limited to adsorption phenomena.    Despite the scientific interest of this result, however, it also reveals a fatal flaw for Li-air batteries, since thick uniform films of lithium peroxide block electron transfer and lead to inefficient battery cycling~\cite{viswanathan2011electrical}.

\begin{figure}
\includegraphics[width=3.5in]{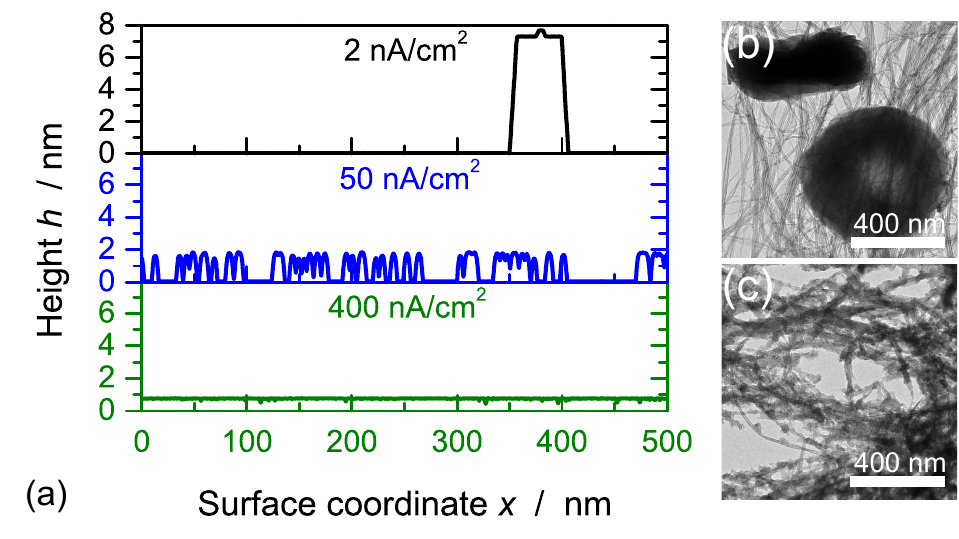}
\caption{ Control of pattern formation by electro-autocatalysis for lithium peroxide electrodeposition in Li-air battery cathodes~\cite{horstmann2013}, predicted by the same general theory.  (a) Phase-field simulations of surface height evolution driven by generalized Butler-Volmer kinetics, which capture the observed voltage behavior (not shown) and morphological transitions with increasing current.  Homogeneous nucleation and growth of islands at very low currents leads to the experimentally observed disk-like particles of Li$_2$O$_2$ shown in (b) on carbon nanotube current collectors. At larger currents, the growth becomes more random and ultimately uniform layer-by-layer above the predicted critical current, as observed in (c).
[Adapted from \citet{horstmann2013}.] }
\end{figure}

\section{ Outlook }

\subsection{  Solid State Ionics}

{\it Li-ion Batteries.} This work opens the possibility of designing interfaces of intercalation materials to control phase separation~\cite{bazant2013,lim2016origin}, as well as structural phase transitions at electrochemical interfaces~\cite{kornyshev1995phase}.    A key goal to improve the rate capability and lifetime of Li-ion batteries is to suppress phase separation during ion insertion, which can be the rate limiting step for both discharging and re-charging of the battery, corresponding to ion insertion at the cathode or anode, respectively.     In contrast, ion extraction at the opposite electrode tends to require less overpotential and might not be affected as much by phase separation. At either electrode, surface phase separation reduces the active area (to that of the exposed phase boundary) and causes degradation via mechanical deformation and  side reactions, such as Li metal plating at the anode during recharging (leading to capacity fade and safety hazards), which become favored once surface ion concentrations reach stable phases.

The theory provides some guidance for surface modification of the active materials to achieve these goals. 
For  generalized Butler-Volmer kinetics (\ref{eq:BV}), the solo-autocatalytic rate is related to the transition state activity coefficient, $S\propto \frac{\partial \gamma_{\ddag}^{-1}}{\partial c}$, which can be altered by blocking sites to reduce the configurational entropy or depositing coatings that cause attractive or repulsive  forces with intercalated ions to adjust the enthalpy. 
These effects may play a role in the observed (but poorly understood) rate-enhancing effects of phosphate or other surficial glass films on LFP and other cathode materials for Li-ion batteries~\cite{kang2009,sun2011}.   

These ideas can be coupled with existing strategies to alter surface chemistry.  For example, 
Park et al.~\cite{park2012enhanced} showed that anion surface modification of LFP by nitrogen or sulfur adsorption improves the insertion rate capability, which they  attributed  lowering of the barrier for lithium ion insertion ($\mu_\ddag=k_BT\change{\ln}\gamma_\ddag$) by stabilizing the under-coordinated Fe$^{2+}$/Fe$^{3+}$ redox couple at the surface.  This chemical bonding effect should be stronger at low lithium concentrations (as in their {\it ab initio} calculations~\cite{park2012enhanced}) due to the lower conductivity of FePO$_4$ limiting access of electrons to the redox site.  In that case, our theory predicts that if the exchange current were \change{preferentially} enhanced at low concentrations, then the reaction would become more solo-autoinhibitory across the spinodal region, \change{thus} further suppressing phase separation and contributing to the observed rate enhancement.

\change{
{\it Resistive Switching Memory. } Electro-autocatalysis may also find applications in the forming of redox-based resistive random access memories (ReRAM)~\cite{waser2009redox}, which are non-volatile, low-power alternatives to today's flash memory.  Promising examples include Valence Change Memories (VCM), based on the controlled dielectric breakdown of transition metal oxide thin films~\cite{waser2007nanoionics}. In the forming cycle at high voltage, metal interstitials or oxygen vacancies undergo compositional instabilities to form conducting filaments of valence-changed metal cations, which are then used to reversibly short circuit the film as a means of information storage. In thick films of perovskite titanates, the forming step has been observed as a bulk fingering instability of the ``virtual cathode" of condensed oxygen vacancies~\cite{waser1990dc,havel2014electroforming},  but in ultra-thin films, Faradaic surface reactions may play a more dominant role. By tuning the solo-autocatalytic electron transfer  rate at the cathode, e.g. by modifying the surface charge and local cation valence as above, it may be possible to control the most unstable wavelength of the instability during forming, and thus the size and spacing of the conducting filaments. 
}

{\it Hydrogen storage.} These effects are not limited to electrodes but also apply to the intercalation materials  for neutral species.  Hydrogen insertion in silicon or palladium nanoparticles has been explored for hydrogen storage and also arises in catalytic materials. Binary phase separation in PdH nanoparticles has been observed {\it in situ} and manipulated via the hydrogen gas pressure~\cite{baldi2014situ,tang2011observations,ulvestad2015avalanching}. It would be interesting to study the response to sudden, large gas pressure steps  to see if phase separation can be suppressed, leading to faster, more uniform intercalation.  Similar surface modification strategies could also be used to manipulate driven autocatalysis.

\subsection{ Electrokinetics at Liquid Interfaces  }
 
{\it Electrovariable Nanoplasmonics. } A major theme of this Discussion is electrovariable optics, based on the reversible deposition of plasmonic nanoparticles at immiscible liquid interfaces driven by electric fields~\cite{edel2016fundamentals,edel2013self}.   The theory of electrovariable nanoplasmonics focuses on the effective trapping potential in the normal direction  to the interface \cite{flatte2008understanding} and includes a thermodynamic model for nanoparticle adsorption and deposition kinetics\cite{flatte2010electrovariable}. The model focuses on the repulsive electrostatic forces between adsorbed nanoparticles, which have the same charge and polarization in the applied field, and thus predicts a stable uniform monolayer, amenable to fast switching.

\begin{figure}
\begin{center}
\includegraphics[width=3in]{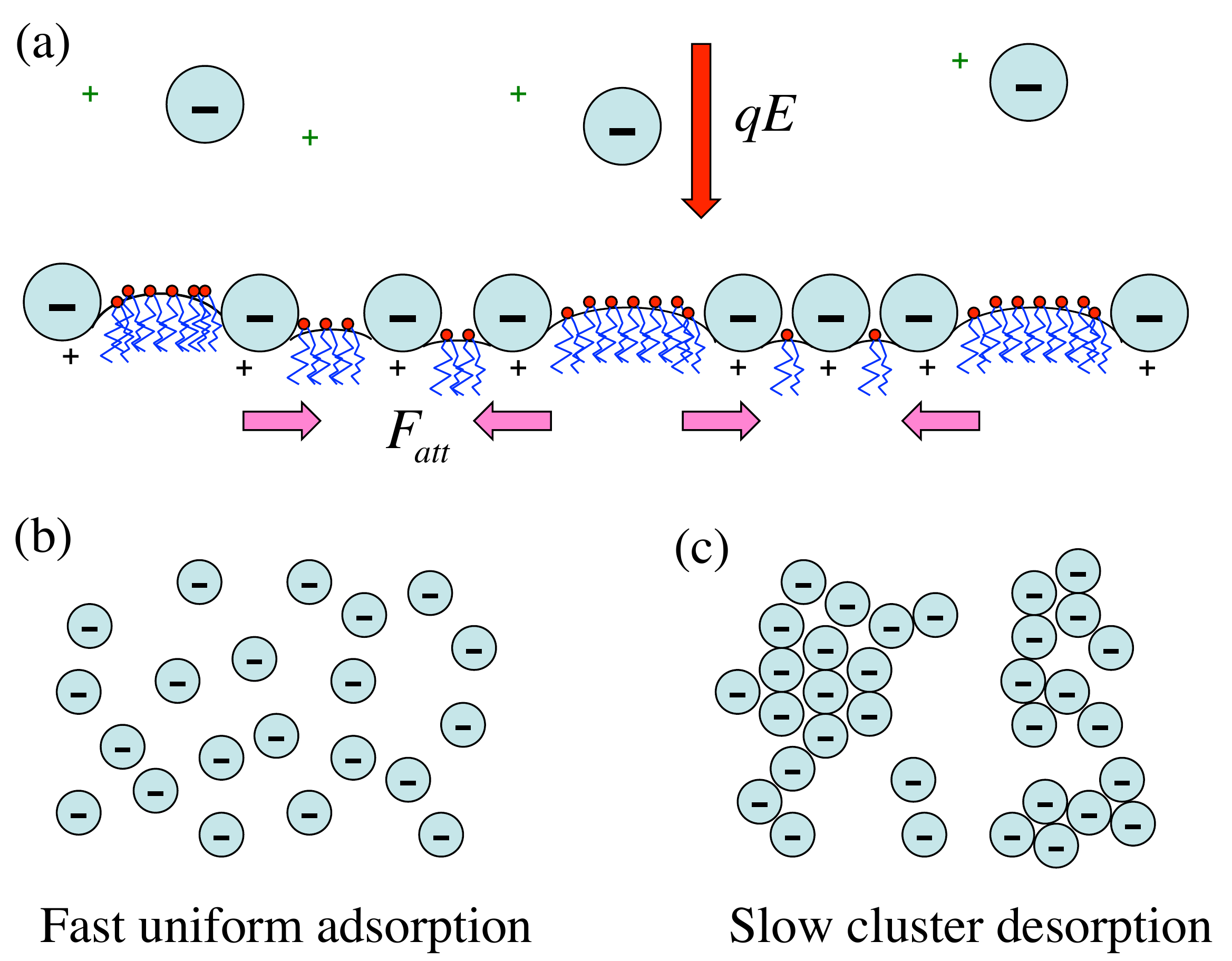}
\end{center}
\caption{ Application of the theory to electrovariable optics with plasmonic nanoparticles adsorbing on an immiscible liquid interface. (a) Attractive lateral forces $F_{att}$ (pink) can result from depletion forces of surfactants or capillary forces mediated by nano-menisci and compete with adsorption driven by the normal electric field. (b) For auto-inhibitory insertion, phase separation can be suppressed (b) above a critical rate, leading to fast, uniform insertion, but in that case, (c) the reverse autocatalytic extraction would destabilize the monolayer and promote phase separation, leading to slow interfacial extraction of clusters. 
\label{fig:mirrors} }
\end{figure}

Although electrostatic repulsion may dominate, there are other strong forces at the nanoscale that could lead to lateral attraction and thermodynamic instability of a nanoparticle monolayer, as shown in Fig. ~\ref{fig:mirrors}(a).  Depending on contract angles, electrostatic forces  and transient geometrical constraints, attractive capillary ``dimple" forces can be very strong for nano-menisci and could lead to clustering (the ``cheerios effect"\cite{vella2005cheerios}).  Moreover,  attractive entropic depletion forces can be tuned by adding surfactants to the system, which would cause inter-particle attraction to reduce the excluded volume for surfactants.

In the presence of attractive lateral forces, the response to an applied electric field becomes more interesting. If the system tends to phase separate into clusters at the interface, then the present theory predicts that electro-catalytic adsorption reactions, e.g. obeying generalized Butler-Volmer kinetics (\ref{eq:BV}) with regular solution thermodynamics (\ref{eq:regsol}),  would stabilize the interfacial monolayer during deposition (Fig.~\ref{fig:mirrors}(b)) and destabilize it during desorption (Fig.~\ref{fig:mirrors}(c)), if the adsorption  reaction is auto-inhibitory, and vice versa, if it is autocatalytic. Stable uniform deposition should proceed more quickly than unstable cluster dissolution, due to the larger active area~\cite{bazant2013}. This prediction should be tested experimentally and any patterns characterized, especially with enhanced  attractive lateral forces.  Since clustering transitions on the interface affect optical properties, it may be possible to exploit these phenomena in new device designs.  For example, switching between a clustered state and uniform coverage without significant mass transfer from the bulk solution could enable faster switching with tolerable resolution.

{\it Electrophoretic Displays. }
Similar issues arises  in electrophoretic displays or ``electronic paper", where colloidal pigments or particles are shuttled between transparent electrodes in liquid-droplet pixels~\cite{comiskey1998electrophoretic}. It is well known that colloidal dispersion stability is important in the bulk liquid, but there is also ordering on the surface that can interfere with device operation~\cite{murau1978understanding}. This clustering contributes to the inability of electronic paper to switch fast enough to enable the playing of movies, and perhaps it could be better understood or even controlled using the ideas in this paper.

{\it  \change{I}onic Liquids and Solids. } 
Room temperature ionic liquids exhibit complex charge oscillations at electrified interfaces~\cite{fedorov2014}. To some extent, these phenomena can be understood in terms of ion crowding~\cite{kornyshev2007} and over-screening due to strong Coulomb correlations~\cite{bazant2011,jiang2014dynamics}, but recent models have also included additional short-range forces that improve predictions of double-layer capacitance~\cite{goodwin2017mean} and promote the ``phase separation" of like-charged domains~\cite{gavish2016theory}, in the hope of explaining long-range charge oscillations~\cite{smith2016electrostatic,gebbie2017long} and other patterns~\cite{yochelis2015coupling,fedorov2014}.   

The treatment of attractive short-range forces and lattice repulsion in these models~\cite{gavish2016theory,goodwin2017mean} is similar to Cahn's regular solution model for binary solid alloys~\cite{cahn1958}, also considered here for solid-state ion intercalation~\cite{bazant2013}, Eq. (\ref{eq:regsol}), although electrostatic interactions are also considered.  As such, the principles of electroautocatalysis and clustering  described here might be useful in understanding the switching dynamics and ordering of ions in applications to electric double layer capacitors.  Moreover, the coupling between double layer structure and Faradaic reactions may be  important in understanding the large electrochemical window of ``solvent-in-salt" ionic-liquid-based electrolytes for rechargeable batteries~\cite{suo2013new,suo2015water}. 

It should be emphasized that our analysis here does not explicitly consider electric fields from diffuse charge or other long-range forces. The stability analysis is based on a homogeneous free energy for short range forces plus a gradient correction, and the focus is on electrochemical reactions in neutral electrolytes with negligible double layer effects.  Past mean-field theories of solid electrolytes~\cite{kornyshev1981} and defect dynamics~\cite{meyer2001advances} accounting for interactions between space charge and Butler-Volmer kinetics have not found any unusual effects on phase separation, so care must be take in applying our results to charged systems. 

\change{
\subsection{  Patterns Driven by Electron Transfer  Reactions }
}
\change{
{\it  Electron Transfer in Solution. }  Electron transfer reactions between donor and acceptor atoms have mostly been studied by chemists at the molecular level without considering how quantum mechanical effects might influence macroscopic reaction-diffusion phenomena.   }  An intriguing new prediction of our theory is the destabilizing effect of negative differential reaction resistance, which is the defining characteristic of the ``inverted region" of Marcus kinetics for outer-sphere electron transfer~\cite{kuznetsov_book,bard_book,bazant2013}.  It is interesting to note that it took several decades \change{
after the pioneering work of Marcus~\cite{marcus1956,marcus1993}, until Miller, Calcaterra and Closs~\cite{miller1984intramolecular,closs1988intramolecular} managed to observe the predicted} effect of exothermocity ($\Delta_r G$) on the kinetics of {\it intra}molecular electron transfer, after many inconclusive studies on {\it inter-}molecular electron transfer.  

Our results suggest that thermodynamic instability of reactive electrolytes in the inverted region of inter-molecular electron transfer could have played a role in these experimental  challenges, 
\change{ due to the coupling of reactions with rapid density fluctuations.  In order to test this prediction,   it would be interesting to revisit the original experiments~\cite{miller1984intramolecular,closs1988intramolecular}, by measuring density fluctuations of the redox species (e.g. by x-ray or neutron adsorption spectroscopy) following a pulse of solvated electrons in a reactive liquid electrolyte (e.g. biphenyl anions in acceptor organic solvents).  Using combinations of intra- and inter-molecular electron transfer, it may be possible to use our theory to control the instability to achieve new types of nanoscale patterns for materials synthesis or actuation. }

\change{
{\it Electron Transfer at Electrodes. } 
Perhaps for similar reasons, it took just as long after Marcus' theory of electron transfer on electrodes~\cite{marcus1965} before it was first verified experimentally by  
~\citet{chidsey1991}, again for intra-molecular electron transfer, across self-assembled monolayers.  Recently,  the first evidence of Marcus-Hush-Chidsey kinetics was reported for solid-solid electron transfer in porous electrodes of Li-ion batteries, where the rate-limiting step was attributed to electron transfer between the iron redox site in LFP and the carbon coating of the active particles~\cite{bai2014}.  While these experiments showed the importance of coupling electron transfer reactions with compositional dynamics, however, they did not probe the effects of negative differential resistance, since integration over the Fermi distribution of electrons eliminates the inverted region for a metallic electrode~\cite{zeng2014MHC,zeng2015simple}.
}

\change{
Last year,  the Marcus inverted region was observed for the first time on a semiconductor photo-electrode\cite{ihly2016tuning}, which, according to our theory, could lead to novel instabilities and pattern formation in photo-electrochemical devices. Yet again, the experiments involved  intra-molecular electron transfer, from single-walled carbon nanotubes to acceptor molecules in fullerene derivatives. Besides photo-electrochemistry and photovoltaics, functionalized carbon nanotubes and graphene sheets are also increasingly used for dynamical processes, such as thermopower waves for electrical energy generation~\cite{liu2016electrical}, where our theory could shed light on the stability of heat and mass transfer coupled with electron transfer reactions. 
}

\change{
\subsection{  Electro-autocatalytic Control of Microstructure }
}

{\it Hydration and Precipitation of Cement Paste.}  One of the most important examples of electrochemical pattern formation is the hydration of tricalcium silicate (the main mineral component of portland cement)  and precipitation of calcium-silica-hydrate (C-S-H)  gels \change{ to form cement paste}~\cite{chen2004solubility,thomas2009influence}. This multi-step reaction is known to be autocatalytic~\cite{thomas2009influence}.  \change{ M}icrostructural evolution, leading to the unique strength of cement paste, proceeds by spinodal decomposition of precipitating C-S-H particles, as shown in recent molecular simulations~\cite{ioannidou2016mesoscale,ioannidou2016crucial}. 

The hardening of cement paste is a natural candidate for continuum modeling with our reactive phase-field model, Eqs. (\ref{eq:genreact}) and (\ref{eq:reactdiff}).  Our general stability analysis \change{ may help explain} how electroautocatalytic precipitation reactions drive spinodal decomposition to determine the final microstructure.  
\change{ The theory may also provide insights into how the paste microstructure could be controlled by varying the initial mixture composition or by applying electrical current during curing, since electric fields are already known to induce microstructural changes in hardened cement~\cite{castellote2002synchrotron}.  More generally, models of driven precipitation may find applications in other electrochemical systems, such as aqueous Li-air batteries~\cite{horstmann2013precipitation}. 
} 

\change{
{\it Electrodeposition.}  We have already discussed how electro-autocatalysis enables morphological control of lithium peroxide electrodeposits~\cite{horstmann2013}, and there are many other possible applications in electro-deposition/dissolution.  For example, nanostructured redox-polymer electrodes for super capacitors have been made by simultaneous electropolymerization of pyrrole and electro-precipitation of polyvinylferrocene~\cite{tian2015electrochemically}, and the microstructure depends on the applied current and electrolyte composition.  
}

\change{
{\it Corrosion.}
A more direct application of our theory arises in the corrosion of binary metallic solids~\cite{erlebacher2001evolution,li2014dealloying}. The de-alloying of a Ag/Au solid solution by selective dissolution of the more electrochemically active metal (Ag) can leave behind three-dimensional nano-porous structures of the more noble metal (Au), which result from modulation of the corrosion rate by surface spinodal decomposition. It was found that simulations based on the regular-solution Cahn-Hilliard equation  could reproduce the experimental microstructure only for a certain choice of the concentration-dependent exchange current~\cite{erlebacher2001evolution}, $I_0\propto e^{-\tilde{c}/c^*}$, rather than the usual assumption of mass action kinetics, proportional to the silver adatom concentration, $I_0 \propto (1-\tilde{c})$.  In hindsight, this is yet another phenomenon of phase-separation control by electro-autocatalysis, which could be tailored to achieve a desired pore size guided by our theory.  
}

\ 

\subsection{ Control of Phase Separation in Biology }

{\it Bacteria and Active Matter.}  Here we have focused on open chemical systems driven by externally controlled reactions, but there are other types of driving work that could fit into our general theoretical framework.  For example, active Brownian suspensions of swimming particles, such as {\it E Coli} bacteria, exert ``swimming stress" on their surroundings~\cite{takatori2014swim}, which leads to phase separation that can be described by a nonequilibrium free energy~\cite{takatori2014swim,speck2014effective,takatori2015towards,takatori2016forces,speck2016stochastic}.  
Equation  (\ref{eq:modglan}) implies that, at least near a nonequilbrium steady state, active diffusion is generally destabilizing, while passive diffusion is stabilizing. Introducing reactions among active swimmers with environmental chemicals could provide an interesting means of tuning their pattern formation. 

{\it Protein Phase Separation in Cells.}  Over the past decade, there has been growing appreciation of phase separation in biology~\cite{hyman2014liquid,brangwynne2015polymer}, stimulated by the discovery of liquid-liquid protein phase separation inside the cytoplasm of embryonic cells, leading to  cell division~\cite{brangwynne2009germline}.   Although it is known that reactions such as RNA/protein binding play a role in controlling phase separation~\cite{zhang2015rna}, most experiments and models have focused either on applied protein concentration gradients without reactions~\cite{lee2013spatial} or on the evolution of already formed droplets~\cite{zwicker2016growth}, regulated by autocatalytic reactions~\cite{zwicker2014centrosomes}, including suppression of Ostwald ripening~\cite{zwicker2015suppression}. 

The present theory could be useful in understanding how driven autocatalytic reactions can stabilize the homogeneous mixture or trigger the onset phase separation and control the nascent patterns the lead to liquid organelles. The general notion of pattern formation by chemically driven phase separation has a long history in biology relating to the origins of life~\cite{zwicker2016growth}. This possibility also fascinated Prigogine~\cite{prigogine1982being}  and motivated much of his own work in nonequilibrium thermodynamics~\cite{kondepudi_book}. 


\section*{ Acknowledgments } 
This work \change{ began during a sabbatical leave at Stanford University} supported by the Global Climate and Energy Project and by the US Department of Energy, Basic Energy Sciences through the SUNCAT Center for Interface Science and Catalysis. The author is grateful to Peng Bai \change{ and Yiyang Li} for help with the figures and insights from battery simulations, Dimitrios Fraggedakis for checking the calculations and noting the second term in Eq. (\ref{eq:Lcstab}), and David Zwicker, Sho Takatori, Thomas Petersen, \change{Albert Tianxiang Liu and Dimitrios} for valuable references.





\bibliography{elec46} 
\bibliographystyle{rsc} 

\end{document}